\documentclass[10pt,journal,compsoc]{IEEEtran}
\ifCLASSOPTIONcompsoc
  \usepackage[nocompress]{cite}
\else
  \usepackage{cite}
\fi
\usepackage{amsmath,amsfonts,bm}
\usepackage{algorithmic}
\usepackage{algorithm}
\usepackage{array}
\usepackage[caption=false,font=footnotesize,labelfont=rm,textfont=rm]{subfig}
\usepackage{textcomp}
\usepackage{stfloats}
\usepackage{url}
\usepackage{verbatim}
\usepackage{graphicx}
\usepackage{cite}
\usepackage{color}
\usepackage{amssymb}
\usepackage{diagbox}
\usepackage{epsfig}
\usepackage{setspace}
\ifCLASSINFOpdf
\else
\fi
\begin{document}
%
\title{SD-AETO: Service Deployment Enabled Adaptive Edge Task Offloading in MEC}
%
%
%
%

\author{Liangjun Song, Gang~Sun,~\IEEEmembership{Member,~IEEE,}
        Hongfang~Yu,~\IEEEmembership{Member,~IEEE,}
        and~Mohsen~Guizani,~\IEEEmembership{Fellow,~IEEE}
\IEEEcompsocitemizethanks{\IEEEcompsocthanksitem This research was partially supported by the National Key Research and Development Program of China (2019YFB1802800), PCL Future Greater- Bay Area Network Facilities for Large-scale Experiments and Applications (PCL2018KP001).\protect\\
L. Song is with the Key Laboratory of Optical Fiber Sensing and Commu- nications, Ministry of Education, University of Electronic Science and Tech- nology of China, Chengdu 6111731, China (e-mail: liangjunsong1231@163.com).\protect\\
G. Sun is with the Key Laboratory of Optical Fiber Sensing and Com- munications, Ministry of Education, University of Electronic Science and Technology of China, Chengdu 6111731, China, and is also with the Agile and Intelligent Computing Key Laboratory of Sichuan Province, Chengdu, 6111731, China (e-mail: gangsun@uestc.edu.cn).\protect\\
H. Yu is with the Key Laboratory of Optical Fiber Sensing and Communica- tions, Ministry of Education, University of Electronic Science and Technology of China, Chengdu 6111731, China, and is also with the Pengcheng Labora- tory, Shenzhen 518000, China (e-mail: yuhf@uestc.edu.cn).\protect\\
M. Guizani is with the Machine Learning Department, Mohamed Bin Zayed University Of Artificial Intelligence (MBZUAI), Abu Dhabi, UAE (e-mail: mguizani@ieee.org).}
\thanks{Manuscript received April 19, 2005; revised August 26, 2015.}}

%
%

\markboth{Journal of \LaTeX\ Class Files,~Vol.~14, No.~8, August~2015}%
{Shell \MakeLowercase{\textit{et al.}}: SD-AETO: Service Deployment Enabled Adaptive Edge Task Offloading in MEC}
%



\IEEEtitleabstractindextext{%
\begin{abstract}
In recent years, edge computing, as an important pillar for future networks, has been developed rapidly. Task offloading is a key part of edge computing that can provide computing resources for resource-constrained devices to run computing-intensive applications, accelerate computing speed and save energy.
An efficient and feasible task offloading scheme can not only greatly improve the quality of experience (QoE) but also provide strong support and assistance for 5G/B5G networks, the industrial Internet of Things (IIoT), computing networks and so on. 
To achieve these goals, this paper proposes an adaptive edge task offloading scheme assisted by service deployment (SD-AETO) focusing on the optimization of the energy utilization ratio (EUR) and the processing latency.
In the pre-implementation stage of the SD-AETO scheme, a service deployment scheme is invoked to assist with task offloading considering each service's popularity. The optimal service deployment scheme is obtained by using the approximate deployment graph (AD-graph).
Furthermore, a task scheduling and queue offloading design procedure is proposed to complete the SD-AETO scheme based on the task priority. The task priority is generated by the corresponding service popularity and task offloading direction.
Finally, we analyze our SD-AETO scheme and compare it with related approaches, and the results show that our scheme has a higher edge offloading rate and lower resource consumption for massive task scenarios in the edge network.
\end{abstract}

}

\maketitle

\IEEEdisplaynontitleabstractindextext

%
\IEEEpeerreviewmaketitle

\IEEEraisesectionheading{\section{Introduction}\label{sec:introduction}}

%
%
%
%
\IEEEPARstart{I}{n} recent years, the rapid development of wireless communication technology and the explosive growth of network users have accelerated the era of big data and low-latency networks. In this era, based on the Internet of Everything (IoE) and aiming at achieving the real-time and high-speed transmission of massive data, the data processing capacity of cloud computing centers has begun to transfer to the network edge, introducing a new computing paradigm, that is, multiaccess edge computing (MEC) \cite{9052677,9444339,9606607,9442819}. MEC sinks the traditional cloud computing capability and provides computing, storage, communication, acceleration, artificial intelligence and big data processing capabilities at the edge of the network close to the user equipment (UE) \cite{9466151}. In addition, MEC provides an open deployment platform for third-party service applications to save return bandwidth and reduce service latency \cite{9123504}.

With ETSI's reference architecture and the mature concept of MEC, 5G MEC is realized by a 5G core network (5GC), edge computing platform and UEs to meet the requirements of billing, legal interception, mobility management and quality of service (QoS) in edge scenarios \cite{9596610}. Therefore, as a 5G native function, MEC will help perform application localization, content distribution and computing marginalization, which is highly consistent with the concept of expanding vertical industry and service-oriented networks in the future, and MEC has therefore become the critical technology and foundation for the development of 5G/B5G, the industrial Internet of Things (IoT) and computing networks \cite{9488278,9426913}. In addition, in contrast to traditional central cloud computing technology, MEC integrates telecommunications and Internet Technology (IT) services, provides cloud computing capability for UEs and other network devices at the edge of the wireless access network, and can reduce the latency of computing, storage, processing and access of UEs \cite{8852687}. It also has the advantages of proximity, high bandwidth, real-time monitoring of network information, location awareness, mobile support and high security \cite{9509282}. These advantages enable MEC to be applied in vertical industries with variable latency requirements, such as the tactile internet, remote surgery, industrial automation, intelligent transportation and smart grids.

The main technologies of MEC systems include low-latency network processing, task offloading and wireless data caching \cite{9274481}. These technologies are the premise and foundation of MEC systems in achieving real-time computing processing, localization of data processing and efficient information interaction. As a core part of the MEC system, task offloading is an important means by which the MEC system achieves real-time processing of terminal services. The main process of task offloading includes three parts: offloading decisions, strategy implementation and result return. Task offloading can also be understood as "computing power redistribution", where the three basic components of "redistribution" are \emph{i)} decision-making in allocation, that is, determining how to reallocate computing power; \emph{ii)} designing the computing resource scheduling algorithm for MEC; and \emph{iii)} performing mobility management for UEs \cite{9488291,9180064}. However, not all MEC systems can undertake offloading tasks in the network. One of the reasons is that an edge server has limited resources and cannot handle all tasks while meeting latency requirements \cite{9452108}. Another reason is that offloading can only be performed when the processing equipment in the network can provide corresponding services for the task \cite{9132684}. Therefore, service deployment or caching is a prerequisite for task offloading and will directly affect performance, thus affecting the UEs' quality of experience (QoE) \cite{9290124}. In the offloading decision step, the traditional offloading mode and constraints can no longer meet the requirements of today's edge tasks. With the development of vertical industry, the functions of tasks have become pluralistic, resulting in different levels of importance and priority for each task. For instance, the task of predicting traffic accidents must be more important than the task of measuring air humidity, so the former should have a higher priority when performing task offloading.

To put it crudely, task offloading needs to consider two steps, service deployment and offloading decisions, both of which are indispensable. Determining how to meet the performance requirements of edge tasks with priority requirements and how to balance resource utilization and computing latency for the MEC edge service rate is an urgent problem to be considered and discussed in research on edge task offloading.

In view of the above considerations, we propose an adaptive edge task offloading scheme assisted by service deployment (SD-AETO).

The main contributions of this paper are as follows:
\begin{itemize}
  \item{We propose the SD-AETO scheme. On the premise of meeting the hit rate requirements of edge services and the priority constraints of offloading tasks, this method takes the service deployment scheme between edge servers as the preprocessing step and carries out the edge task offloading scheme between UEs in the IoT that can adapt to different edge task requirements.}
  \item{We model the edge layer as a graph structure. By solving the quota problem of Steiner trees on a graph composed of edge servers storing different services, we reduce the cache space of edge services and reduce the energy consumption of edge servers.}
  \item{We consider the priority constraint problem of edge computing offloading tasks at the UEs and servers and integrate the priority constraint problems of these two objects into an integrated constraint problem, which greatly reduces the decision-making time and the complexity of task offloading while meeting the priority constraints of different tasks.}
  \item{We conduct extensive simulation experiments to validate and evaluate our proposed SD-AETO scheme.} 
\end{itemize}

The remainder of this paper is organized as follows: Section II reviews and summarizes the related work. Section III introduces the problem formula and the framework of the SD-AETO scheme. Section IV describes the details of the pretreatment method for offloading objects. Section V presents the task offloading execution process. Section VI presents the simulation results and compares our algorithms with other information sensing algorithms, followed by the conclusion in Section VII.

\section{Related work}
\subsection{Research on edge service deployment}
Edge service deployment integrates traditional service deployment methods and mechanisms into the edge computing infrastructure. By moving memory storage closer to UEs, edge service deployment reduces the pressure on the network and improves content delivery. An efficient service deployment scheme can promote the optimization of edge task offloading performance, so it has become the focus of many researchers.

Francesc Guim et al. \cite{d9612603} proposed a resource management and configuration scheme in an intelligent dynamic scenario based on service level objectives (SLOs) to guarantee edge services. This scheme takes autonomous cycle management on the edge platform as the center and achieves the goal of dynamic resource scheduling of multitenant services. 
To solve the problem of multiservice deployment in the MEC network environment, Lu Zhao et al. \cite{d9452122} described how to minimize the utilization of edge resources while meeting the quality of service required by edge users. They used a heuristic algorithm based on priority to realize the edge deployment mode of distributed control and maximize the overall service quality of users under the condition of comprehensive consideration of service sharing and communication interference.
Tian Wang et al. \cite{d9543585} proposed an edge intelligent service deployment algorithm. This algorithm uses the theory of simulated annealing to adjust the particle position in the particle swarm optimization algorithm to find the global optimal solution.

The above schemes studied the problem of service deployment in depth, but there is no interface combined with task offloading, which requires further exploration.

\subsection{Research on edge task offloading}
Task offloading in edge computing is an important research problem. Different time delay and offloading schemes of mobile devices have a great impact on the completion of tasks and mobile device energy consumption. This makes it difficult and crucial to formulate a reasonable task offloading strategy to adapt to dynamic changes in the environment. 

Yuqing Li et al. \cite{t8798727} proposed an online learning offloading mechanism assisted by an accumulated trust value. This mechanism decouples a series of two-way offloading problems and uses Lyapunov optimization to explore the long-term spatiotemporal optimality of the system. However, this scheme can only choose between robustness and stability and cannot achieve the overall optimal goal of the task offloading scheme.
In 2021, reference \cite{t9375490} proposed a computational offloading scheme aimed at maximizing rewards, which tends to offload tasks with high power demand and high reward. It uses a traffic graph to model the task allocation problem and proposes two offloading subalgorithms based on task segmentation to solve the offloading problem.
Because environmental factors such as transmission efficiency and resource constraints in the edge network are complex and changeable, edge devices will be unstable when processing tasks. To solve this problem, Xiangjie Kong et al. \cite{t9165923} proposed an optimization framework for an edge cooperative network to improve the performance of task offloading.
In addition, Chunmei Ma et al. \cite{t9344808} incorporated the concept of edge computing offloading into the Internet of Vehicles environment and used parked vehicles as virtual MEC servers to solve the problem of limited computing resources. On this basis, a local task scheduling strategy was proposed to further improve the performance of task offloading.
Qing Li et al. \cite{t9123603} proposed an algorithm to save energy by reducing the QoS in the offloading process of edge computing in 2022. This algorithm uses a statistical calculation model and statistical transmission model to quantify the compatibility between the QoS and offloading strategy and achieves reduced energy consumption.

Although the above task offloading strategies optimize performance from multiple perspectives, the services cached on the MEC servers, which are required by the tasks as a prerequisite for the success of task offloading, are rarely mentioned, let alone explored and studied accordingly.

\subsection{Comprehensive research on joint consideration of\\ service deployment and task offloading}
Scholars have conducted much research on edge service deployment and task offloading. However, considering the integrity of edge networks, no single study can determine the optimal scheme of edge computing offloading.

According to the different load capacities and cache services of different MEC servers, Lei Wang et al. \cite{c9279254} used the Sinkhorn matrix to solve the calculation offloading problem of edge tasks to dynamically balance the calculation load between MEC servers and provide offloading services for users with the fewest communication hops.
Meng Qin et al. \cite{c9165797} studied the problem of multiple radio access technologies simultaneously offloading delay-sensitive tasks. This problem has great significance for the computing offload and service caching of massive tasks. In the algorithm proposed in reference \cite{c9165797}, with strict delay requirements and residual energy requirements as constraints, the goal of minimizing delay and energy overhead is achieved through a distributed optimization model.
Reference \cite{c9599706} studied the problem of request offloading and collaborative service caching between 5G MEC and cloud data centers under the condition of meeting user QoS and base station resource capacity requirements. It proposed a two-level optimization framework and obtained the best dynamic service deployment rules and reward distribution method.

Although the above schemes take into account the relationship between task offloading and service deployment, they all focus on task offloading in the research process and lack exploration and analysis of the correlation between them.

In summary, although many scholars have studied service deployment and task offloading in the MEC scenario, only a few related works have fully studied both of them. It is still very difficult to find a method to synchronously ensure optimal deployment resources and edge service rates, and this requires further research and discussion.

\section{Preliminaries and system model description}
In the 5G and industrial big data environment, network devices generate a large number of latency-sensitive tasks and medium or large tasks that cannot be processed locally through sensors, and they offload these tasks to the MEC server of the network for computing and processing. Reference \cite{9155233} showed that 45\% of the data in the IoT need to be stored, processed and executed at the edge layer. Different network devices have dedicated network roles with various function requirements, so they will request various services from and offload various tasks to the MEC server. 

\begin{figure}[!t]
  \centering
  \includegraphics[width=2.5in]{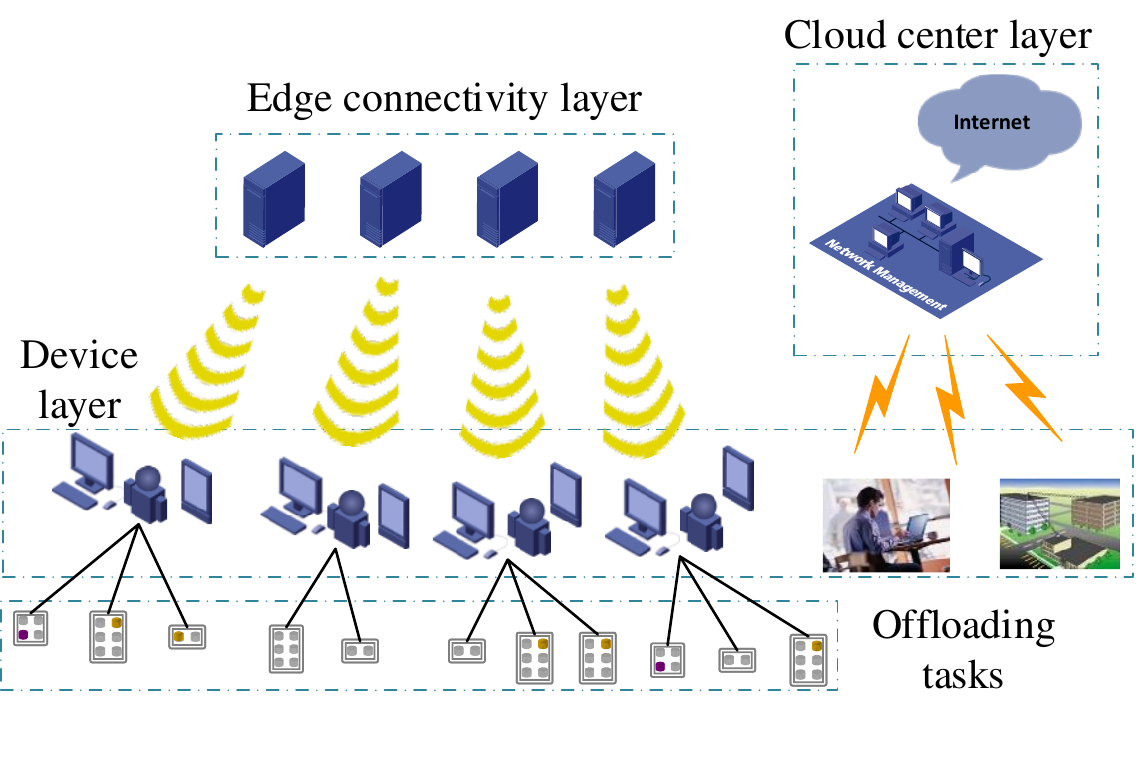}
  \caption{System model.}
  \label{system}
\end{figure}

To perform adaptive edge task offloading while meeting edge service rate and task priority requirements, we take the "cloud-edge-device" three-tier architecture as the system model of the proposed SD-AETO scheme. Figure \ref{system} shows this three-tier architecture.
First, the cloud center layer refers to large network processing servers such as cloud servers and data centers in the core network, and the number of servers in the cloud center layer is $\mathbb{C}$. These servers have extremely high computing, storage and communication capacity. However, due to the long distance from the UEs, the cloud servers cannot meet the latency requirement of the UEs, so they are unable to perform latency-sensitive tasks. 
Second, the edge connectivity layer is composed of $M$ MEC servers deployed at the edge of the network. In contrast to the cloud server, the MEC server is placed on the base station close to the UEs and has limited computing, storage and communication capacity; it provides high-reliability and low-latency services for the UEs. 
Third, the device layer is composed of the UEs in the edge network. The UEs offload the tasks that they cannot handle to the server and directly obtain the processing results after calculation and processing by the server. We assume that the set of UEs is $\mathbb{U}$ and the number of UEs covered by the \emph{i-th} MEC is $U_i$. Therefore, the size of set $\mathbb{U}$ can be represented as follows:
\begin{equation}
\label{size_UEs}
\lvert\mathbb{U} \rvert = \sum_{i=1}^M U_i
\end{equation}

The set of UEs can be obtained from Equation \eqref{set_UEs}:
\begin{equation}
\label{set_UEs}
\mathbb{U} = \left\{u_1, u_2, ..., u_{\lvert\mathbb{U} \rvert}\right\}
\end{equation}

The set of tasks $\mathbb{F}$ generated by all UEs is shown as Equation \eqref{set_tasks}. 
\begin{equation}
  \label{set_tasks}
  \mathbb{F} = \left\{task_1, task_2, ..., task_{\mathbb{U}}\right\}
\end{equation}

where \emph{$task_i$} represents the set of tasks generated by the \emph{i-th} UE.

Under this system model, this paper first implements a service deployment strategy based on service popularity with the goal of minimizing storage space in the edge connectivity layer and then completes the task offloading method based on task popularity with the goal of minimizing execution time.

To clarify our SD-AETO scheme, the following assumptions are stated:
\begin{itemize}
  \item{The arrival of the UEs' task offloading request and the new service obey a Bernoulli stochastic process with probabilities $B_u$ and $B_s$, respectively.}
  \item{There is no priority constraint between different UEs.}
  \item{Due to the high computing power of cloud servers, the computing delay on them is ignored in this paper.}
  \item{In view of the proximity characteristics of MEC, this paper ignores the propagation delay while tasks are offloaded to MEC from the UEs.} 
  \item {The MEC in the network is always available, and there are no outages.}
\end{itemize}

In short, the whole SD-AETO scheme described above can be divided into two stages: the task offloading preprocessing stage and the task offloading execution stage. In more detail, the task offloading preprocessing stage is divided into two parts: service popularity design and service deployment in the edge connectivity layer. In addition, considering the priority constraint of the tasks, the task offloading execution stage can be divided into the task scheduling part, centered on priority setting, and the task offloading queue design part. We will give the details in Section IV and Section V, respectively.
The key notation in this paper is shown in Table \ref{notation}.

\begin{table}[!t]
  \caption{A practical instance of an offload matrix\label{notation}}
  \centering
  \begin{tabular}{|c||l|}
  \hline
  $M$& The number of MEC servers in edge connectivity layer.\\
  \hline
  $\mathbb{U}$& The set of UEs.\\
  \hline
  $\mathbb{F}$& The set of tasks.\\
  \hline
  $B_u$& The task arrival rate.\\
  \hline
  $B_s$& The service arrival rate.\\
  \hline
  $\mathbb{S}$& The existing service set in the network.\\
  \hline
  $P_{s_i}$& The service popularity.\\
  \hline
  $A$& The number of approximate microservices.\\
  \hline
  $R$& The sum of the given quota rewards.\\
  \hline
  $T_{total}$&The maximum processing latency.\\
  \hline
  $b$& The symbol of task beginning processing time.\\
  \hline
  $l$& The symbol of task completion processing time.\\
  \hline
  $\varLambda $& The subtask offloading matrix.\\
  \hline
  $\mathcal{E} $& The candidate queue.\\
  \hline
  $\mathcal{Q} $& The completed offloading queue.\\
  \hline
  \end{tabular}
\end{table}

\section{SD-AETO scheme pre-implementation: Service deployment}
To keep the task offloading process proceeding and improve the edge offloading rate (the edge service rate) of the edge connectivity layer as much as possible, it is necessary to deploy current hot-spot services on the MEC server. 
However, the storage resources on the MEC server are relatively limited. 
Therefore, when deploying hot-spot services, we should not only consider the impact of service popularity on the edge offloading rate but also minimize the occupation of storage resources on the MEC server. 
Next, we will introduce service popularity design and edge service deployment in detail.

\subsection{Service's popularity design}
Let the existing service set in the network be $\mathbb{S}$ and the set of services deployed in the edge connectivity layer be \bm{$S_M$}.
If the total number of services in the network is $N$ and $C$ services among them are deployed on MEC servers, $\mathbb{S}$ and \bm{$S_M$} can be defined as follows:
\begin{equation}
  \label{set_all_services}
  \mathbb{S} = \left\{s_1, s_2, ..., s_N\right\}
\end{equation}
\begin{equation}
  \label{set_MEC_services}
  \mathbf{S_M} = \left\{s_1, s_2, ..., s_C\right\}
\end{equation}

Then, let the service popularity corresponding to each service be $P(s_i)$, which also represents the probability that the task sent by the UE exactly corresponds to service $s_i$. The relationship among service popularities in the network is:
\begin{equation}
  \label{sum_popularity}
  \sum_{i=1}^Ns_i = 1
\end{equation}

Thus, the edge service hit rate $P^{hit}$ can be derived from Formula \eqref{hit_rate}.
\begin{equation}
  \label{hit_rate}
  P^{hit} = \sum\nolimits_{s_i\in \mathbf{S_M}}P(s_i)
\end{equation}

At the time at which a new service $s_{new}$ with service probability $P(s_{new})$ enters the network, we update the set of services $\mathbb{S}$ to \bm{$S_{new}$}, as shown in Equation \eqref{new_set_services}.
\begin{equation}
  \label{new_set_services}
  \mathbf{S_{new}} = \left\{s_1, s_2, ..., s_N, s_{new}\right\}
\end{equation}

To ensure that the sum of all service probabilities is $1$ after the new service $s_{new}$ is pushed into the network, the service probabilities other than $s_{new}$ are updated to $P_{new}(s_i)$ in accordance with Formula \eqref{old_new_probability}.
\begin{equation}
  \label{old_new_probability}
  P_{new}(s_i) = P(s_i)(1-P(s_{new}))
\end{equation}
where
\begin{equation}
  \label{new_sum_1}
  P_{new}(s_1) + P_{new}(s_2) + ... + P_{new}(s_N) + P(s_{new}) = 1
\end{equation}

Therefore, the edge service hit rate when $s_{new}$ is not cached in the edge connectivity layer but is cached in the cloud center layer is shown as follows:
\begin{equation}
  \begin{aligned}
    \label{null_hit_rate}
    P_{null}^{hit} &= \frac{\sum_{s_i\in \mathbf{S_M}}P_{new}(s_i)}{\sum_{i=1}^NP_{new}(s_i)+P(s_{new})}\\
                   &= P^{hit}(1-P(s_{new}))
  \end{aligned}
\end{equation} 

Formula \eqref{cache_hit_rate} shows the edge service hit rate after $s_{new}$ is cached on the MEC servers with the updated set $S_{new}$.
\begin{equation}
  \begin{aligned}
    \label{cache_hit_rate}
    P_{cache}^{hit} &= \frac{\sum_{s_i\in \mathbf{S_{new}}}P_{new}(s_i)}{\sum_{i=1}^NP_{new}(s_i)+P(s_{new})}\\
                    &= P_{null}^{hit} + P(s_{new})
  \end{aligned}
\end{equation} 

Next, we consider a more common scenario.

Because hot-spot services change over time, we define the service popularity update time slot $\zeta = \left\{0,1,2,...,\Gamma-1\right\}$.
Suppose $\alpha$ new services are pushed in the network, and the popularities of $\aleph$ new services are less than the deployment threshold $\Delta$, where $0 \leq \Delta \leq 1$.
Let the set of new probabilities $\mathbb{G}$ be as follows:
\begin{equation}
  \label{range_propularity}
  \mathbb{G} = \left\{g_1, g_2, ..., g_{\alpha-\aleph}\right\}
\end{equation}
For $i = 1, 2, ..., \alpha-\aleph$, $\aleph \leq g_i \leq 1$ represents the probability of the corresponding service.

Then, the distribution function of $P_i(s_{new})$ can be obtained according to Formula \eqref{distribution_function}.
\begin{equation}
  \begin{aligned}
    \label{distribution_function}
    \mathcal{F}(P_i(s_{new})) &= \frac{I(P_i(s_{new})\in \mathbb{G})}{\alpha-\aleph}B_s \\
                              &+ I(P_i(s_{new})=0)(1-B_s)
  \end{aligned}
\end{equation}
where $I(\cdot)$ is the indicator function. $P_i(s_{new})=0$ represents that the corresponding service $s_{new}$ is not pushed in the edge connectivity layer.

Through the above service popularity design process, the services in the network are configured with their own service popularities. Next, the edge service deployment operation is performed based on these popularities.

\subsection{Edge service deployment}
The service deployment method of the edge connectivity layer has a great impact on the edge offloading rate, but MEC servers with limited capacity cannot place all the required services. 
Therefore, the purpose of implementing the edge service deployment process in the SD-AETO scheme is to minimize the edge service cache space while meeting the requirement of the edge offloading rate. 
It should be emphasized that in this paper, the edge offloading rate is equivalent to the hit rate of edge cache services and thus equivalent to the sum of the service popularities of the services deployed on the MEC servers.

In view of the correlation between time and geography, there is service deployment redundancy between MEC servers.
To reduce the deployment space of edge services, this part stores each service on the edge server in the form of a microservice and performs a de-redundancy operation on the microservices deployed on each MEC.

Let service $s_i = \left\{s_i^1, s_i^2, ..., s_i^k\right\}$ in the network be a set of a series of microservices, where $k$ is the total number of microservices divided by $s_i$. The storage space occupied by the deployment services before de-redundancy, $\varphi$, is shown as follows:
\begin{equation}
  \label{space}
  \varphi = \sum\nolimits_{i\in {[1,N_{new}]}}|s_i|
\end{equation}
where $N_{new}$ represents the total number of services deployed within the current network.

Arranging the services in the current network in descending order of service popularity, we can obtain the service set \bm{$S_D$} by Equation \eqref{descending_service}.
\begin{equation}
  \label{descending_service}
  \mathbf{S_D} = \left\{s_1^{new}, s_2^{new}, ..., s_{N_{new}}^{new}\right\}
\end{equation}

We can obtain the storage space occupied by the deployment services at the edge connectivity layer after de-redundancy, $\varphi (\Omega )$, according to Formula \eqref{de_edge_space}.
\begin{equation}
  \label{de_edge_space}
  \varphi (\Omega ) = |s_1^{new}\cup s_2^{new}\cup ... \cup s_w^{new}|
\end{equation}
where $\Omega$ is the set of services cached on the MEC servers and $w$ represents the total number of services deployed at the edge connectivity layer.

Similar to $\varphi (\Omega )$, let $\Psi$ represent the set of services cached in the cloud center layer. The storage space occupied by the services pushed in the network at the cloud center layer is obtained from Formula \eqref{de_cloud_space}.
\begin{equation}
  \label{de_cloud_space}
  \varphi(\Psi) = \varphi (\Omega ) + \sum\nolimits_{i\in {[w+1, N_{new}]}}|s_i|
\end{equation}

Let the maximum storage capacity of the MEC server $M_i(i=1,2,..,M)$ be $\xi_i $ and the minimum requirement of the edge offloading rate be $B$. The following \emph{deployment optimization problem} can be obtained:
\begin{subequations}
  \label{service_optimization}
  \begin{align}
    min  \qquad  &\varphi (\Omega ) \label{target} \\
    s.t. \qquad  &\varphi (\Omega )\leq \sum_{i=1}^M\xi_i,\label{capacity}\\
                 &\sum_{{s_i}\in \Omega} P_{new}(s_i)\geq B\label{offloading_rate}\\
    var. \qquad  &\Omega \subseteq \Psi \label{variable}
  \end{align}
\end{subequations}

Inequality \eqref{capacity} is the capacity constraint condition, which means that the storage space occupied by the services deployed in the edge connectivity layer cannot exceed the total capacity of the MEC servers.
Constraint condition \eqref{offloading_rate} ensures that the edge offloading rate will not be lower than $B$.
Formula \eqref{variable} represents that all the services that need to be deployed on MEC servers have been pushed into the cloud center layer.

Although it seems intuitive, it is actually a great challenge to solve this problem since $\varphi (\Omega)$ is the result of the de-redundancy of various services and the quantification of $\varphi (\Omega)$ has an open form.

\textbf{Claim 1: The \emph{deployment optimization problem} is NP-hard.}

\emph{proof:} 
To prove the above claim, we will show that the above optimization problem is NP hard even if there is no redundancy between MEC servers by analogy to the knapsack problem. By extension, we consider a knapsack problem with a set of $n$ items, as follows:
\begin{equation}
  \label{item_knapsack}
  \mathbb{I} = \left\{1, 2, ..., n\right\}
\end{equation}

Suppose the \emph{i-th} item's value is $v_i$ and $w_i$ represents its volume. The knapsack problem is to obtain an item subset $\mathbb{J}\subseteq \mathbb{I} $ to maximize the value of the items in the knapsack; that is:
\begin{subequations}
  \label{knapsack}
  \begin{align}
    max  \qquad  &\sum_{i\in\mathbb{J} }v_i \\
    s.t. \qquad  &\sum_{i\in\mathbb{J} }w_i\leq W
  \end{align}
\end{subequations}
where $W$ is the knapsack volume, $v_i\geq 0$, and $w_i\geq 0$.

Next, we reduce the knapsack problem to the \emph{deployment optimization problem}.

First, we create a set of services as follows:
\begin{equation}
  \label{service_knapsack}
  \mathcal{S} = \left\{s_1, s_2, ..., s_n\right\}
\end{equation} 

Second, service $s_i$ is composed of $\mathcal{W}_i$ microservices, so $\mathcal{W}_i$ can be regarded as the volume. 

Third, the value of a service is represented by the service probability $P(s_i)$.

Since there is no redundancy between MEC servers, the goal of the optimization problem proposed is simplified as Formula \eqref{goal_simplified}.
\begin{equation}
  \label{goal_simplified}
  \varphi (\Omega ) = \sum_{{i}\in \Omega}\mathcal{W}_i
\end{equation}

Therefore, if $\sum_{i=1}^M\xi_i=W$, the solution of the \emph{deployment optimization problem} can also solve the knapsack problem. 

Thus, we can conclude that the \emph{deployment optimization problem} is NP-hard. 

Q.E.D.

Therefore, to obtain the edge service deployment strategy, we obtain the optimal solution of the deployment strategy in the edge connectivity layer by abstracting the MEC servers and the services cached on them as elements in the approximate deployment graph (AD-graph).

\subsubsection{AD-graph}
\begin{figure}[!t]
  \centering
  \includegraphics[width=2.5in]{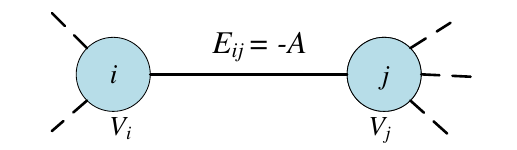}
  \caption{Construction of approximate graph.}
  \label{approximate_graph}
\end{figure}

As shown in Figure \ref{approximate_graph}, we construct an approximation graph $G = (V, E) $. 
The nodes in the graph represent the MEC servers in the edge connectivity layer, while the edges represent that there are approximate microservices between the two associated nodes. Naturally, if no approximate microservice exists between two MEC servers, there is no edge between the corresponding nodes. 
Let the node weight $V_i$ represent the number of microservices deployed on the homologous MEC server; the edge weight $E_{ij} = -A$ represents the negative value of the number of approximate microservices deployed on the two associated nodes.

This graph can effectively estimate the space requirements for deploying services after de-redundancy. From then on, the above spatial constraints of optimization problem \eqref{service_optimization} can be represented by finding a subgraph of the AD-graph.

\subsubsection{Quota problem for Steiner tree}\label{Quota problem of Steiner tree}
To integrate the optimization objectives \eqref{target} into the approximate graph, a quota reward parameter $r_i$ is allocated to each node of the approximate graph, which is equivalent to the sum of service popularities deployed on the corresponding MEC server, as shown in Formula \eqref{quota_parameter}. 
\begin{equation}
  \label{quota_parameter}
  r_i = \sum\nolimits_{s_i \prec M_i}P(s_i) 
\end{equation}

\begin{figure}[!t]
  \centering
  \includegraphics[width=2.5in]{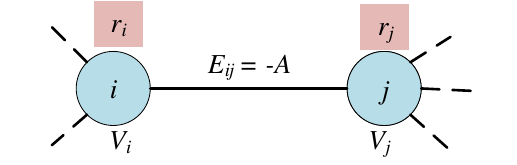}
  \caption{The Steiner tree problem with quota rewards.}
  \label{quota_graph}
\end{figure}

Figure \ref{quota_graph} represents the Steiner tree  problem with quota rewards.

Thus, the quota problem of Steiner trees can be solved on the basis of Figure \ref{quota_graph}. That is, a subgraph $G^{\prime}=(V^{\prime}, E^{\prime})$ with the minimum sum of node weights and edge weights can be obtained under the condition of satisfying the sum $R$ of the given quota rewards. The sum $R$ of quota rewards is the required edge offloading rate, which can be shown as follows:
\begin{equation}
  \label{quota_reward}
  R = \sum_{i=1}^{|V^{\prime}|}r_i
\end{equation}
where $V^{\prime}$ represents the set of nodes after de-redundancy and the sum of node weights and edge weights of the subgraph $G^{\prime}$ is the de-redundant storage space $\varphi (\Omega )$ occupied by the deployed services in the edge connectivity layer.

In other words, we should find a microservice deployment strategy with the highest service popularities under the condition of meeting the capacity constraints to achieve the goal of occupying the minimum capacity space while obtaining the maximum edge offloading rate, equal to the sum $R$ of the given quota reward.

\subsubsection{Transformation to a k-minimum spanning tree (k-MST) problem}
\begin{figure}[!t]
  \centering
  \includegraphics[width=2.5in]{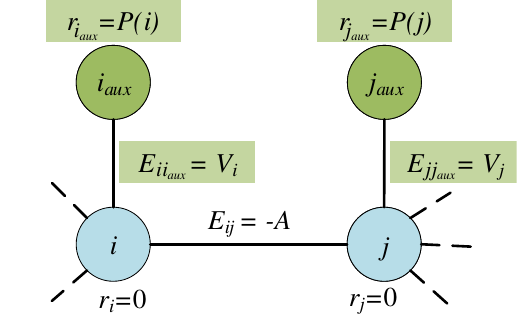}
  \caption{Transformation to a k-minimum spanning tree (k-MST) problem.}
  \label{k-MST}
\end{figure}

As shown in Figure \ref{k-MST}, to simplify the process of generating subgraphs and improve computational efficiency, we convert the AD-graph containing node quota rewards mentioned in section \ref{Quota problem of Steiner tree} into a k-minimum spanning tree (k-MST) problem to perform subsequent calculations. Specifically, the graph with edge weights and node costs is transformed into a graph that contains edge weights only.  

First, we change the popularity $P(v) = r_i$ of vertex $v$ of the AD graph with quota rewards to zero. 

Second, we add an affiliated adjacent node $v_{aux}$ on the vertex $v$ mentioned above, and the quota reward $P(v_{aux})$ of the adjacent node $v_{aux}$ is as follows:
\begin{equation}
  \label{adjacent_node_quota}
  P(v_{aux}) = P(v) = r_i
\end{equation}

Finally, Formula \eqref{add_edge_weight} represents the weight of the edge of the associated adjacent node $v_{aux}$ and the original vertex $v$, which is set to the node weight of the original vertex $v$.
\begin{equation}
  \label{add_edge_weight}
  E_{vv_{aux}} = V_i
\end{equation}

\subsubsection{Equivalent update}
To make the transformed k-MST problem solve the quota problem of the Steiner tree equivalently, we update the quota rewards $P(v)$ and $P(v_{aux})$ of the original vertex $v$ and the adjacent node $v_{aux}$ according to Formulas \eqref{quota_old} and \eqref{quota_new}.
\begin{equation}
  \label{quota_old}
  P^{\ast}(v) = 2|V^{\ast}|P(v) + 1
\end{equation}
\begin{equation}
  \label{quota_new}
  P^{\ast}(v_{aux}) = 2|V^{\ast}|P(v_{aux}) + 1
\end{equation}
where $V^{\ast}$ represents the set of nodes of the newly generated graph $G^{\ast} = (V^{\ast}, E^{\ast})$.

In the new graph, the sum of the given quota rewards $R$ is also updated as follows:
\begin{equation}
  \label{quota_reward_new}
  R^{\ast} = 2R|V^{\ast}|
\end{equation}

\begin{figure}[!t]
  \centering
  \includegraphics[width=2.5in]{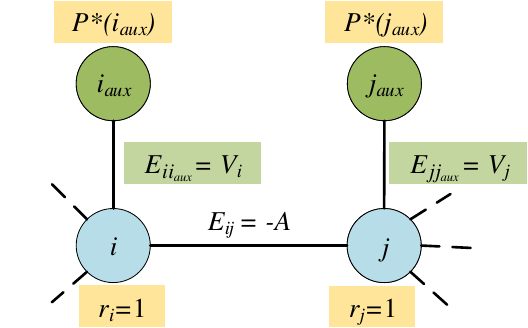}
  \caption{Equivalent update.}
  \label{update}
\end{figure}

The schematic diagram after the equivalent update is given in Figure \ref{update}.

\textbf{Claim 2: The optimal solution to the quota problem with updated vertex quota rewards \bm{$P^{\ast}(v_{aux})$} and \bm{$P^{\ast}(v)$} and the sum of the given quota rewards \bm{$R^{\ast}$} is also optimal for the original quota problem with vertex quota rewards \bm{$P(v_{aux})$} and \bm{$P(v)$} and the sum of the given quota rewards \bm{$R$}.}

\emph{Proof \cite{quota}:}
Let tree $T$ be the optimal solution; we will prove its feasibility and optimality.

Feasibility: Considering any tree $T$ that is feasible under the updated quota reward, we have:
\begin{equation}
  \label{proof_1}
  \sum_{v^{\ast}\in T} \geq R^{\ast} = 2|V^{\ast}|R
\end{equation}

According to the original quota reward, we can rewrite Formula \eqref{proof_1} as follows:
\begin{equation}
  \label{proof_2}
  \sum_{v^{\ast}\in T}P^{\ast}(v_{\ast}) = 2|V^{\ast}|\sum_{v^{\ast}\in T}P(v^{\ast}) + |T| \geq 2|V^{\ast}|R
\end{equation}

According to the characteristics of the spanning tree, when each parameter in the graph is multiplied by a sufficiently large constant, the resulting spanning tree remains unchanged. Next, we can assume that all quota rewards are integers. 
Since tree $T$ spans no more than $|V^{\ast}|$ nodes, Formulas \eqref{proof_3} and \eqref{proof_4} can be obtained under the above hypothetical condition.
\begin{equation}
  \label{proof_3}
  2|V^{\ast}|\sum_{v^{\ast}\in T}P(v^{\ast}) \geq 2|V^{\ast}|R
\end{equation}
\begin{equation}
  \label{proof_4}
  \sum_{v^{\ast}\in T}P(v^{\ast}) \geq R
\end{equation}

Therefore, any tree $T$ that is feasible under the updated quota reward is also feasible for the original quota problem with the vertex quota reward and the sum of the given quota rewards. 

Optimality: Suppose there is another tree $T^{\prime}$ that is better than tree $T$, such that $weight(T^{\prime}) < weight(T)$ with the following condition:
\begin{equation}
  \label{proof_5}
  \sum_{v^{\ast}\in T^{\prime}}P(v^{\ast}) \geq R^{\ast}
\end{equation}

For any span tree, the edge weights are the same since they do not depend on the quota rewards.
Thus, we have a contradiction of the supposition mentioned above.

Q.E.D.

\subsubsection{Star node transformation for the k-MST problem} 
\begin{figure}[!t]
  \centering
  \includegraphics[width=2.5in]{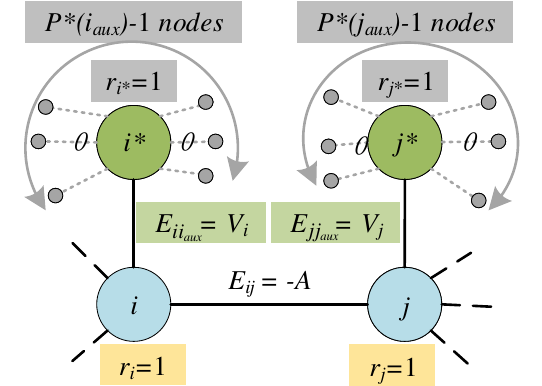}
  \caption{Transformation to star nodes.}
  \label{star_nodes}
\end{figure}

To further process the above graph to make it applicable to the general k-MST algorithm, we convert each vertex $v_{aux}$ into a star node $v^{\ast}$, as shown in Figure \ref{star_nodes}.

Attaching $P^{\ast}(v_{aux})$ subsidiary nodes to each star node $v^{\ast}$, let the weights of the edges connected between these subsidiary nodes and the center of the star node be $0$ and the quota rewards of all vertices be $1$.
Therefore, the optimal solution $T^{\ast}$ for the k-MST problem is also optimal for the original graph under the condition shown in Formula \eqref{k_R}.
\begin{equation}
  \label{k_R}
  k = R^{\ast}
\end{equation}

Thus far, the preprocessing stage of edge task offloading has been completed, and the edge service deployment scheme that minimizes the service storage space has been obtained while meeting the requirements of the edge offloading rate with the consideration of hot-spot service popularities. Algorithm \ref{al_service_deployment} shows the process of service deployment.

\begin{algorithm}[H]
\caption{SD-AETO: service deployment.}
\label{al_service_deployment}
\begin{algorithmic}
\STATE
\STATE {\textbf{INPUT:}} $MEC, \bm{S_D}, \xi_i  $;
\STATE {\textbf{OUTPUT:} optimal solution} $ T^{\ast} $;
\STATE {\textsc{//Construct}} AD-graph: $G = (V, E)$;
\STATE {\textsc{//Transform}} $G$ into $G^{\ast}$;
\STATE \hspace{0.5cm} \textbf{for} each node $i$:
\STATE \hspace{1cm} Update $i$ by $i$ and $i^{\ast}$;
\STATE \hspace{1cm} Assign vertexs' quota rewards according \\
       \hspace{0.5cm} Formulas \eqref{quota_old} and \eqref{quota_new};
\STATE \hspace{0.5cm} \textbf{end for}
\STATE \hspace{0.5cm} \textbf{for} each edge $E_{ij}$:
\STATE \hspace{1cm} Assign edge weights as $-A$ and $E_{vv_{aux}} = V_i$;
\STATE \hspace{0.5cm} \textbf{end for}
\STATE \hspace{0.5cm} Scale total quota reward $R $ to $R^{\ast}$: $R^{\ast} = 2R|V^{\ast}$;
\STATE {\textsc{//Solve quota problem on graph}}: $G^{\ast} = (V^{\ast}, E^{\ast})$;
\STATE \hspace{0.5cm} Let $k = R^{\ast}$;
\STATE \hspace{0.5cm} Solve problem: $ k-MST (G^{\ast}, R^{\ast}) $
\STATE \hspace{0.5cm}\textbf{return} tree$ T^{\ast} $
\end{algorithmic}
\end{algorithm}

\section{SD-AETO scheme implementation: Task scheduling and offloading queue design}
After the above edge task offloading preprocessing stage is completed, MEC servers in the edge connectivity layer can provide corresponding functional support for edge task offloading. Next, the SD-AETO algorithm implementation will be described in detail.
First, because the tasks generated in the "big data era" have their own priorities, the importance differences of various tasks will be taken into account when performing task offloading in this part.
Second, to get the most out of the processing resources of MEC servers and reduce the latency of the offloading process, we divide the task into several subtasks, similar to the case with services.
Finally, we assign an offloading priority to each subtask and propose the method of subtask scheduling and offloading queue design.

\subsection{Modeling of the latency optimization problem}
\begin{figure}[!t]
  \centering
  \includegraphics[width=2.5in]{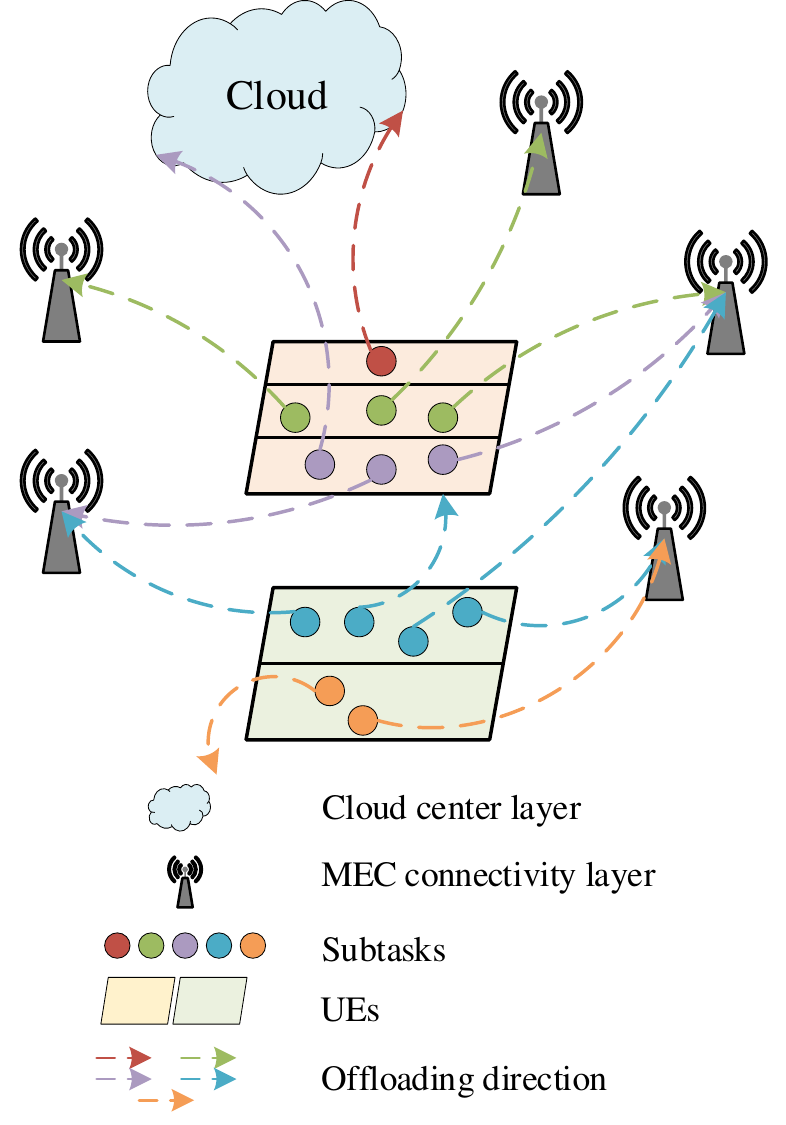}
  \caption{Illustration of tasks offload to different directions.}
  \label{offloading_direction}
\end{figure}

In the existing edge computing network, we can conclude that the task offloading direction of UEs is composed of four parts: local processing, D2D offloading, edge connectivity layer offloading and cloud center layer offloading. 
Note that local processing is not considered in this paper.
Figure \ref{offloading_direction} shows an illustration of the offloading direction in the network.

Let $F$ represent the total number of subtasks in the network.
The objective function of the offloading process in the SD-AETO algorithm aims to minimize the maximum processing latency $T_{total}$ of offloading tasks in the whole network, and it is defined by Formula \eqref{T_total}.
\begin{equation}
  \label{T_total}
  max T_{total} = \mathop{\max}_f t_f^{\eta _f,l}
\end{equation}
where $f \in [1,F]$ represents any subtask in the network, $\eta_f$ represents any task-receiving equipment (such as a MEC server, cloud server, UE, \dots), and $l$ represents the processing completion of subtask $f$.

According to Formula \eqref{T_total}, another meaning of $T_{total}$ is the processing completion time of the last subtask among the received subtasks in the network.

Suppose $t_f^{\eta _f,b}\geq 0$ represents the task beginning processing time, and $t_f^{\eta _f,pr}$ and $t_f^{\eta _f,pd}$ represent the task processing delay and the propagation delay, respectively.
The calculation of the processing completion time $t_f^{\eta _f,l}> 0$ is discussed in the following two cases considering the propagation delay and processing delay.

\emph{case 1:} Offloading to the cloud center layer.
\begin{equation}
  \label{time_cloud}
  t_f^{\eta _f,l} = t_f^{\eta _f,b} + t_f^{\eta _f,pd}
\end{equation}
Considering that the servers in the cloud center layer have extremely high data processing capacity, Formula \eqref{time_cloud} does not consider the processing delay of the tasks in this case. The propagation delay must be calculated because of the long distance between UEs and the cloud center layer. 

\emph{Case 2:} Offloading to the edge connectivity layer or another UE.

In contrast to case 1, Formula \eqref{time_not_cloud} takes the processing delay into account instead of the propagation delay because of the proximity of MEC servers and UEs.
\begin{equation}
  \label{time_not_cloud}
  t_f^{\eta _f,l} = t_f^{\eta _f,b} + t_f^{\eta _f,pr}
\end{equation}
Next, we need to solve the \emph{latency optimization problem} for edge task offloading, shown as follows:
\begin{subequations}
  \label{latency_optimization}
  \begin{align}
    min  \qquad  &max T_{total} \label{offloading_target}\\
    s.t. \qquad  &\sum\nolimits_{H_f}\delta _f = 1 \label{direction}\\
                 &t_{f,u_i}^b - t_{f\ll 1,u_i}^l\geq 0, \label{same_UE}\\
                 &\sum_{i=1}^F\delta_f t_f^{\eta_f,l} \geq \sum_{i=1}^F\delta_{f^{\prime}} t_{f^{\prime}}^{\eta_f,l}, \label{same_direction}\\
                 &\delta_f \in \left\{0,1\right\}, \label{inseparable}\\
                 &t_f^{\eta_f,l} > 0, \label{completion}\\
                 &t_f^{\eta_f,b} \geq 0, \label{beginning}
  \end{align}
\end{subequations}
where $H_f$ is the set of alternative servers and UEs for offloading subtask $f$ and $f\ll 1$ indicates that its offloading priority is lower than that of $f$.

In the latency optimization problem \eqref{latency_optimization}, constraint \eqref{direction} indicates that a subtask cannot be offloaded in multiple directions.
Constraint \eqref{same_UE} reflects the sequence and priority constraints of subtasks sent by the same UE $u_i$. In other words, among the subtasks sent from the same UE, the subtasks with low priority can only be processed after the subtasks with high priority. This means that the offloading direction of subtasks in this constraint can be inconsistent.
Constraint \eqref{same_direction} means that subtasks offloaded in the same direction $n_f$ follow the principle of first come, first served.

Next, the corresponding details of the task scheduling and offloading queue design will be given.

\subsection{Task scheduling based on integration priority}
In time slot $\tau (\tau <<\zeta)$, let the set of tasks generated by UE $u_i$ be $task_{u_i}$, which is shown as follows:
\begin{equation}
  task_{u_i} = \left\{task_{u_i}^1, task_{u_i}^2,...\right\}
\end{equation}
where $u_i\in \mathbb{U} $.

Suppose that each subtask corresponds to a microservice, which means the subtask can only be offloaded to a server or other UE with the microservice cached, and the offloading objects of each subtask include all edge servers, cloud servers and UEs.

To reduce the resource consumption and processing delay caused by considering the task priority constraint, this step sets the offloading priority as a floating point number, where the integer part is the main priority, determining the offloading object of the subtask, and the fractional part is the subordinate priority, determining the offloading queue as described in the next subsection.

\subsubsection{Subtask redefinition}
We reconstruct the task $task_{u_i}$ into subtasks $task_{u_i}^{sub}$ as follows:
\begin{equation}
  \label{re_subtask}
  task_{u_i}^{sub} = \left\{task_{u_i}^{sub,1}, task_{u_i}^{sub,2}, ..., task_{u_i}^{sub,h_i}\right\}
\end{equation}
where $task_{u_i}^{sub,\beta }(\beta \in[1,h_i])$ represents a subtask sent by UE $u_i$.

\subsubsection{Offloading matrix}
According to the SD-AETO algorithm pre-implementation process, a subtask offloading matrix $\varLambda$ with a size of $\sum_{i=1}^{|\mathbb{U} |}h_i\times (M+C+|\mathbb{U}|)$ can be constructed, in which each row represents the offloading objects of subtask $task_{u_i}^{sub,\beta}$ and each column represents the subtasks that can be received by servers and UEs. In addition, the element $\varLambda _{ij}$ represents the latency when the subtask corresponding to the row element $\varLambda_{i\cdot }$ is offloaded to the server or UE corresponding to the column element $\varLambda_{\cdot j }$.
Table \ref{offload_matrix} shows a practical instance of an offload matrix in an edge computing network, where ``-'' indicates that the microservice corresponding to subtask $\varLambda_{i\cdot }$ is not deployed on $\varLambda_{\cdot j }$ during the task offloading preprocessing phase.
Note that the latency of MEC servers and UEs is $t_f^{n_f,pr}$, while $t_f^{n_f,pd}$ represents the latency of cloud servers.
\renewcommand\arraystretch{2}
\begin{table}[!t]
  \caption{A practical instance of an offload matrix\label{offload_matrix}}
  \centering
  \begin{tabular}{|l||c|c|c|c|c|c|}
  \hline
  & $M_1$& $M_2$& $Cloud_1$& $UE_1$& $UE_2$& $UE_3$\\
  \hline
  $task_{u_1}^{sub,1}$& 5& 4& 7& -& -& 8\\
  \hline
  $task_{u_2}^{sub,1}$& -& 8& 10& -& -& -\\
  \hline
  $task_{u_2}^{sub,2}$& 2& -& 6& 5& -& -\\
  \hline
  $task_{u_3}^{sub,1}$& 8& -& 5& -& -& -\\
  \hline
  $task_{u_3}^{sub,2}$& -& 3& 7& 9& 6& -\\
  \hline
  \end{tabular}
\end{table}

\subsubsection{Offloading object sequence}
The offloading object sequence of each subtask $task_{u_i}^{sub,\beta }$ can be obtained from the non-zero elements of the row referred to in matrix $\varLambda$ according to Formula \eqref{object_sequence}.
\begin{equation}
  \label{object_sequence}
  \begin{aligned}
    &\mathbb{O} _{u_i}^{sub,\beta} = \\
    &\left\{M_{i_1}, M_{i_2},...,Cloud_{j_1}, Cloud_{j_2},..., UE_{k_1}, UE_{k_2},...,UE_{k_e}\right\}
  \end{aligned}
\end{equation}

\begin{figure}[!t]
  \centering
  \includegraphics[width=2.5in]{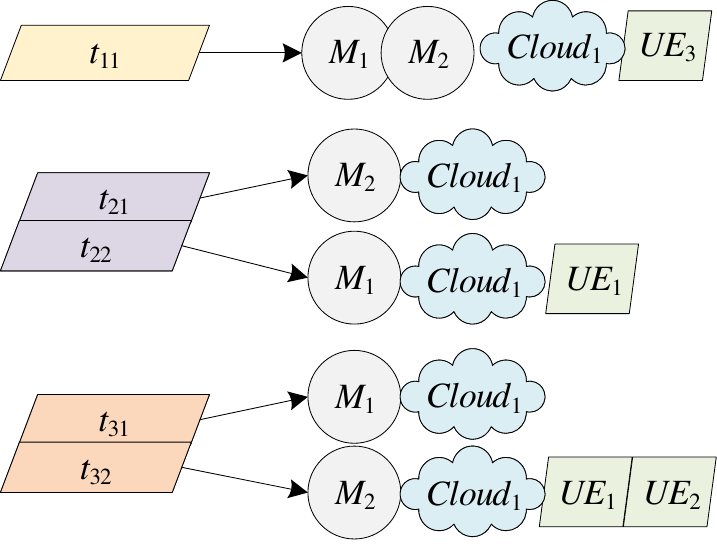}
  \caption{Illustration of offloading object sequence.}
  \label{Offloaded_object_sequence}
\end{figure}

Figure \ref{Offloaded_object_sequence} shows the offloading object sequence of each subtask in matrix $\varLambda$. 
To simplify the illustration, we use $t_{ij}$ to represent subtask $task_{u_i}^{sub,\beta}$.
As an example, the offloading object sequence of $task_{u_3}^{sub,2} = \left\{M_2, Cloud_1, UE_1, UE_2\right\}$ is shown.

\subsubsection{Redefinition of offloading object sequence}
\begin{figure}[!t]
  \centering
  \includegraphics[width=2.5in]{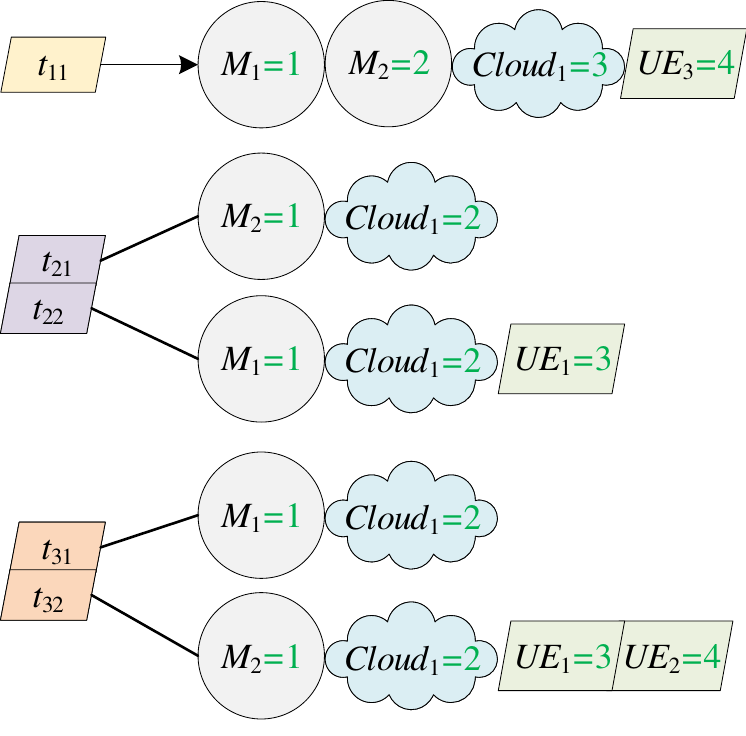}
  \caption{Illustration of redefined offloading object sequence.}
  \label{redefined_object_sequence}
\end{figure}

To save storage space and reduce the computational complexity, we redefine the offloading object sequence as follows:
\begin{equation}
  H_{u_i}^{sub,\beta}(n) = \left\{o_{u_i,1}^{sub,\beta}, o_{u_i,2}^{sub,\beta},...,o_{u_i,n}^{sub,\beta}\right\}
\end{equation}
where $o_{u_i,1}^{sub,\beta} = M_{i_1} = 1$, $o_{u_i,2}^{sub,\beta} = M_{i_2} = 2$, ..., $o_{u_i,n}^{sub,\beta} = UE_{k_e} = n$, and $n$ is the total number of processing devices that can provide offloading services for subtask $task_{u_i}^{sub,\beta}$. 
Figure \ref{redefined_object_sequence} shows the changes in the redefinition of the instance in matrix $\varLambda$.

\subsubsection{Integration priority}
\begin{figure}[!t]
  \centering
  \includegraphics[width=2.5in]{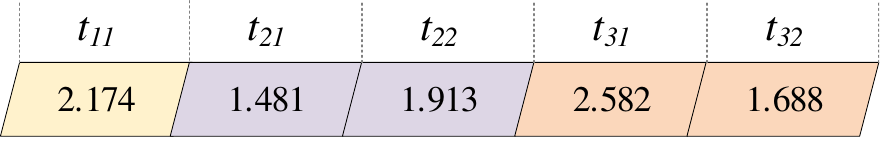}
  \caption{Illustration of integration priorities.}
  \label{integration_priorities}
\end{figure}

Let the serial number $o_{u_i,j}^{sub,\beta} (j\in [i,n])$ with the shortest latency in the redefined offloading object sequence $H_{u_i}^{sub,\beta}(n)$ be the main priority and the popularity of the services to which the microservices belong, corresponding to subtask $task_{u_i}^{sub,\beta}$, be the subordinate priority.
For example, the integration priorities of the subtasks mentioned above are shown in Figure \ref{integration_priorities}.

\begin{spacing}{1.0}
\subsection{Offloading queue design}
\end{spacing}
Although each subtask is assigned its own priority parameter, the offloading directions of each subtask are not exactly the same, and the priority constraints between subtasks should also be satisfied. Therefore, subtasks need to be sorted to obtain the offloading queue to improve the efficiency of the SD-AETO algorithm.
We define two auxiliary empty queues, $\mathcal{E} $ and $\mathcal{Q} $, where $\mathcal{E} $ is the candidate queue and $\mathcal{Q} $ is the completed offloading queue. 
The first step is to select the first subtask from each UE and put it into queue $\mathcal{E} $; i.e., $\mathcal{E} = \left\{t_{11}, t_{21}, t_{31}\right\} $. 
Then, the subtask with the highest priority in $\mathcal{E} $ is put into $\mathcal{Q} $ and removed from $\mathcal{E} $; i.e., $\mathcal{Q} = \left\{t_{31}\right\} $ and $\mathcal{E} = \left\{t_{11}, t_{21}\right\} $. 
Next, the adjacent task of the removed subtasks, i.e., $t_{32}$, needs to be put into $\mathcal{E} $; i.e., $\mathcal{E} = \left\{t_{11}, t_{21}, t_{32}\right\} $. The SD-AETO algorithm repeats the above steps until all subtasks in the time slot $\tau$ are sorted. 
In other words, the steps are repeated until $\mathcal{E} $ is empty and the size of $\mathcal{Q} $ is equal to $F$ in time slot $\tau$. Finally, each subtask can be offloaded to the corresponding server or UE for further processing according to the offloading queue $\mathcal{Q} $. 
Algorithm \ref{al_queue} shows the offloading queue design.

\begin{algorithm}[H]
  \caption{SD-AETO: Offloading queue design.}
  \label{al_queue}
  \begin{algorithmic}
  \STATE
  \STATE {\textbf{INPUT:}} $\mathcal{E}, \mathcal{Q}, parameter(priority), subtasks $;
  \STATE {\textbf{OUTPUT:} offloading queue} $ \mathcal{Q}^{\ast} $;
  \STATE {\textbf{Initialization:} 
  \STATE \hspace{0.5cm} Clear $\mathcal{E}, \mathcal{Q}$;
  \STATE \hspace{0.5cm} Select the first subtask from each UE;
  \STATE \hspace{0.5cm} Put these subtasks into queue:} $\mathcal{E} $;
  \STATE \textbf{do:} $\left\{\right.$
  \STATE \hspace{0.5cm} Select the subtask $t^{\prime}$ with the highest priority in $\mathcal{E} $;
  \STATE \hspace{0.5cm} Put $t^{\prime}$ into $\mathcal{Q}$;
  \STATE \hspace{0.5cm} Remove $t^{\prime}$ from $\mathcal{E} $;
  \STATE \hspace{0.5cm} Put the adjacent task into $\mathcal{E} $;
  \STATE \hspace{0.6cm}$\left.\right\}$ while $|\mathcal{Q}| < F$;
  \STATE \textbf{return} $ \mathcal{Q}^{\ast} $
  \end{algorithmic}
\end{algorithm}

\section{Simulation results and analysis}
In this section, we construct some simulation experiments to verify the validity and applicability of the SD-AETO scheme. 
Specifically, we simulate four groups of parameters, the edge service rate, offloading latency, service deployment redundancy and resource utilization, under different scenarios and conditions to reflect the performance of the SD-AETO scheme.
The simulation results are analyzed by comparing the RBORA scheme proposed in \cite{RBORA} and the IPEO scheme proposed in \cite{IPEO} with the SD-AETO scheme proposed in this paper.

\subsection{Simulation environment and settings}
Under the heterogeneous network (HetNet) architecture, the simulation environment is based on enhanced machine-type communication (eMTC).
This is an emerging concept in 5G based on the 3GPP scenario that evolved from the LTE protocol. The 5G eMTC tailors and optimizes the LTE protocol to make it more suitable for supporting rich and innovative IoT applications. In a massive connection scenario, eMTC can be directly upgraded and deployed based on the LTE network, and it can share a site and antenna feed with the existing LTE base station.
Table \ref{parameter} lists the default simulation parameters used in this paper, if not otherwise specified.
\begin{table}[!t]
  \caption{Parameters of the simulations\label{parameter}}
  \centering
  \begin{tabular}{|c||c|}
  \hline
  \textbf{Parameter}& \textbf{Value}\\
  \hline
  Cell radius& 500m\\
  \hline
  Channel bandwidth& 5 MHz\\
  \hline
  Transmission power& 14 dBm\\
  \hline
  Density of noise power in the channel& $10^{-9}W/Hz$\\
  \hline
  Task acceptance probability  of the UE& 0.95\\
  \hline
  Number of UEs& $[0,100]$\\
  \hline
  Number of MECs& $[0,40]$\\
  \hline
  \end{tabular}
\end{table}

\subsection{Performance comparison}
\begin{figure}[!t]
  \centering
  \includegraphics[width=2.5in]{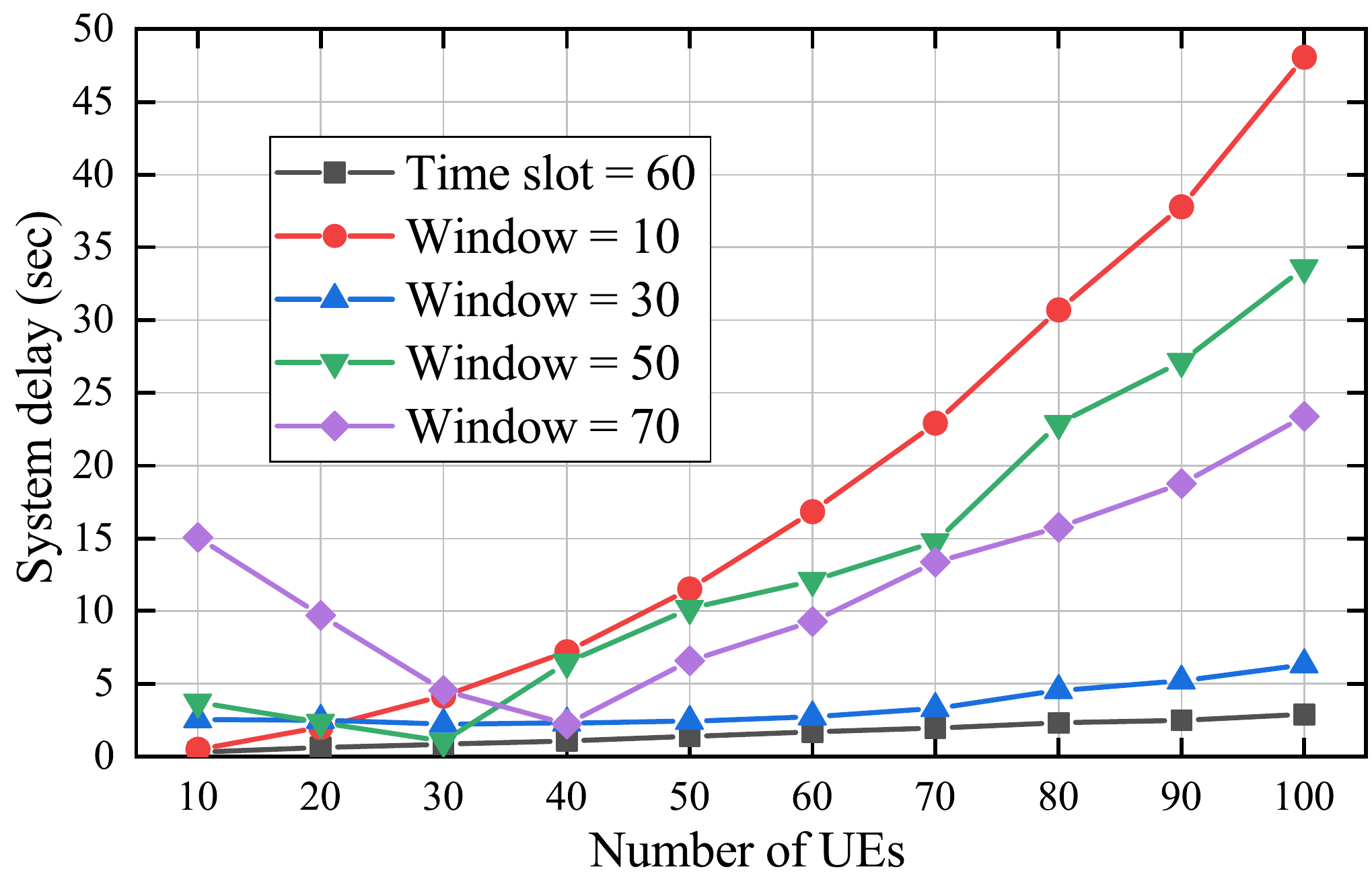}
  \caption{Determination of subtask window.}
  \label{window}
\end{figure}

In this paper, to assign an appropriate offloading priority to each subtask, we need to batch process the subtasks in a short time slot $\tau$, which means that the size of the queue window will directly affect the performance of the SD-AETO scheme.
When the other parameters are fixed, we simulate the system delay under different sizes of windows to set a better parameter for the window size.
As shown in Figure \ref{window}, $window = [10,30,50,70]$ represents the trend in the system delay in the fixed-window mode, and $time slot = 60$ represents the trend in the floating-window mode with a fixed time slot.

When the size of the window $\varepsilon < 30$, i.e., $\varepsilon = 10$, the subtask window can be filled almost without waiting. Therefore, the processing delay of the system will increase with the increase in the number of UEs.
When $\varepsilon \geq  30$, i.e., $\varepsilon = [30, 50, 70]$, the delay curve first increases and then decreases with the increase in UEs.
For further analysis, taking the window size $\varepsilon = 70$ as an example, when there are few UEs, i.e., $UEs \leq 40$, the subtask window needs to wait a long time to be filled to execute the subsequent process each time, and the fewer UEs there are, the longer the waiting time.
When the number of UEs increases, i.e., $UEs \geq 40$, a single window cannot process a whole batch of subtasks, and there will inevitably be subtasks waiting for the next window to process and execute, which will also increase the processing delay of subtasks in the network.

The critical value $UEs^\ast = 40$ can be predicted according to Formula \eqref{critical value}:
\begin{equation}
  \label{critical value}
  (UEs)^\ast = \varepsilon B_u
\end{equation}
where $B_u$ represents the task arrival rate. Note that in the subsequent simulation process, $B_u$ defaults to 0.6 unless specified otherwise.

Furthermore, for the case of $\varepsilon = 30$, we design the floating-window mode to explore better window settings, which is shown as the curve $Time slot = 60$. As seen from Figure \ref{window}, this mode has more advantages than the fixed-window mode, so we will use this floating-window mode in the following simulations.

\begin{figure}[!t]
  \centering
  \includegraphics[width=2.5in]{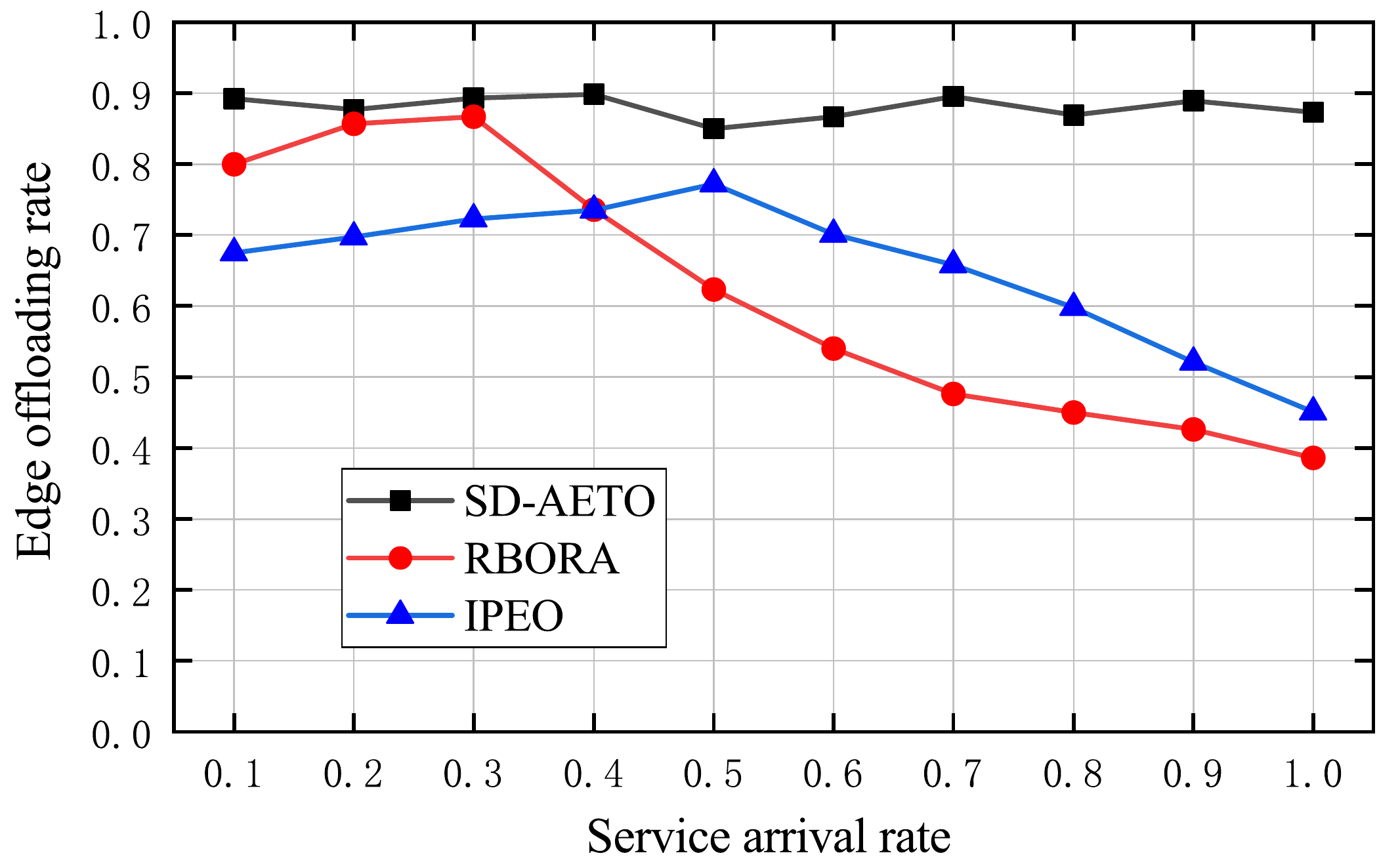}
  \caption{The edge offloading rate with the service arrival rate $B_s$.}
  \label{off_ser}
\end{figure}

Figure \ref{off_ser} shows the trend of the edge offloading rate with the service arrival rate in the SD-AETO scheme and the comparison algorithms IPEO \cite{IPEO} and RBORA \cite{RBORA}.
With the increase in the service arrival rate $B_s$, the edge offloading rate of the SD-AETO scheme proposed in this paper is always higher than those of the comparison algorithms, and it remains at approximately $90\%$. However, the edge offloading rates of both the comparison algorithms first increase and then decrease. The curve of the RBORA scheme reaches the peak value $0.86$ approximately when $B_s = 0.3$, and the peak value $0.79$ of the IPEO scheme is reached at $B_s = 0.5$. 
This brief analysis shows that when the number of services in the network exceeds a certain degree, the comparison algorithms cannot adaptively select a better service deployment scheme, resulting in services with high popularity not being cached by MEC, so the hit rate of edge services is greatly reduced. Compared with the other algorithms, the proposed scheme can better meet the needs of the edge unloading rate and is more suitable for scenarios with a large number of users.

\begin{figure}[!t]
  \centering
  \includegraphics[width=2.5in]{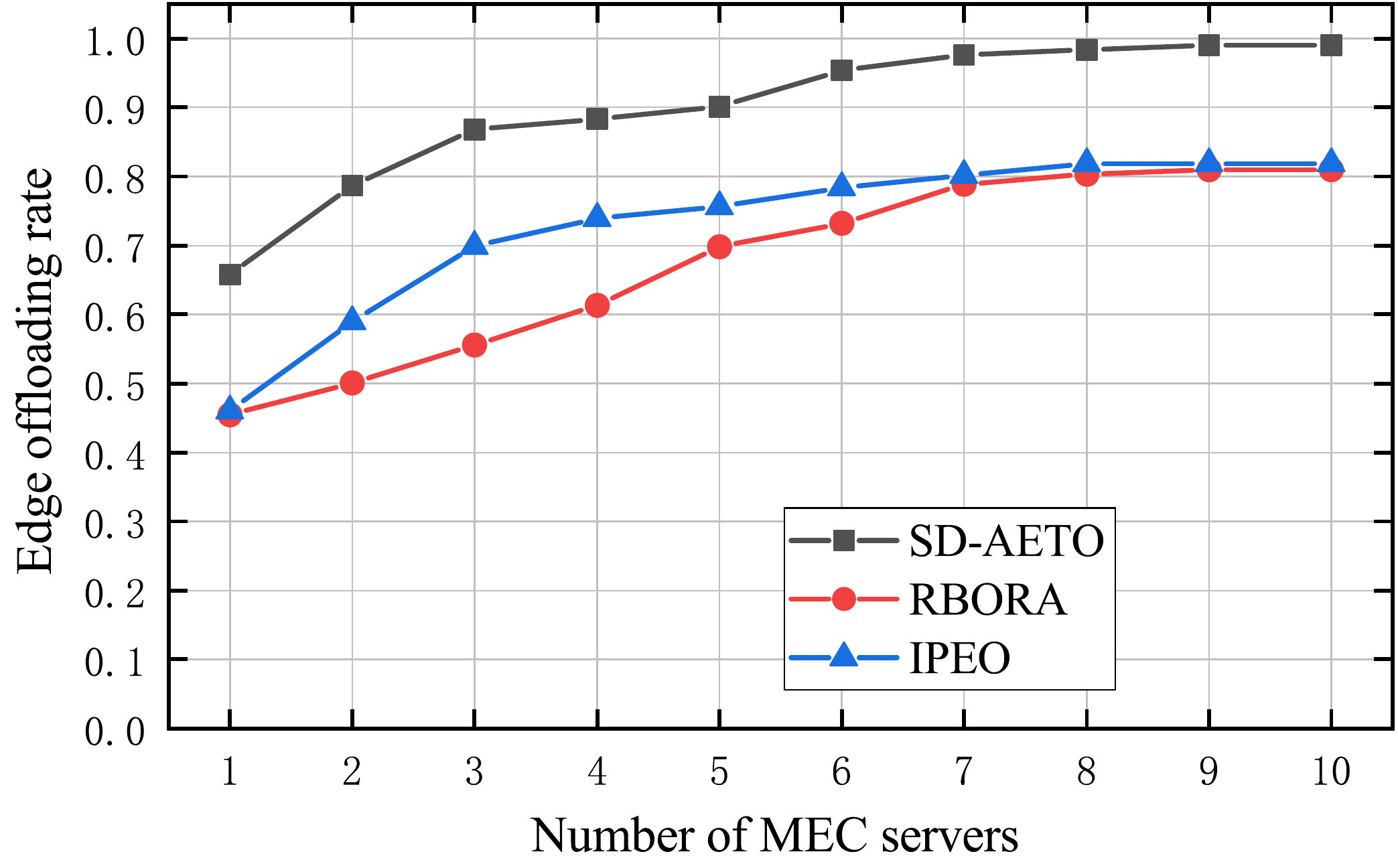}
  \caption{The edge offloading rate with the number of MEC servers.}
  \label{off_mec}
\end{figure}

Figure \ref{off_mec} is a line chart of the change in the edge offloading rate with the number of MEC servers. 
Horizontally, the more MEC servers there are, the more services can be deployed in the edge connectivity layer. Since the edge offloading rate is directly proportional to the number of cached services to a certain extent, the curve in Figure \ref{off_mec} shows an upward trend overall.
Vertically, when the number of MEC servers is equal to $1$, the SD-AETO scheme and comparison algorithm will not perform service deployment at the same time. However, in this paper, the popular services will be repositioned when the time slot $\tau$ is updated. Therefore, the edge offloading rate of our scheme will also be higher than those of the comparison algorithms in this case. 
Furthermore, consider that when the number of MEC servers is greater than $1$, the adaptive service deployment strategy in our scheme will further amplify the above advantage and even achieve the goal of full offloading in the edge environment as long as there are enough MEC servers.
In contrast, in the comparison algorithms, the random deployment method is adopted for the services cached on the MEC servers, and there is no service update mechanism, with the result that even if there are enough MEC servers, they cannot meet the offloading requests of all UEs in the network.

To study the de-redundancy rate of the SD-AETO scheme proposed in this paper, a redundancy degree $\vartheta $ is assigned to measure it as follows:
\begin{equation}
  \label{simulation_redundancy}
  \vartheta = \frac{\varphi (\Omega )-\sum_{T^{\ast}} A^{\ast}}{\varphi (\Omega )}
\end{equation}
where $\sum_{G^{\ast}} A^{\ast}$ represents the quantity of redundant services after deployment.

We simulate the trend of the de-redundancy rate with the change in the service arrival rate and the number of MEC servers in the edge connectivity layer.

\begin{figure}[!t]
  \centering
  \includegraphics[width=2.5in]{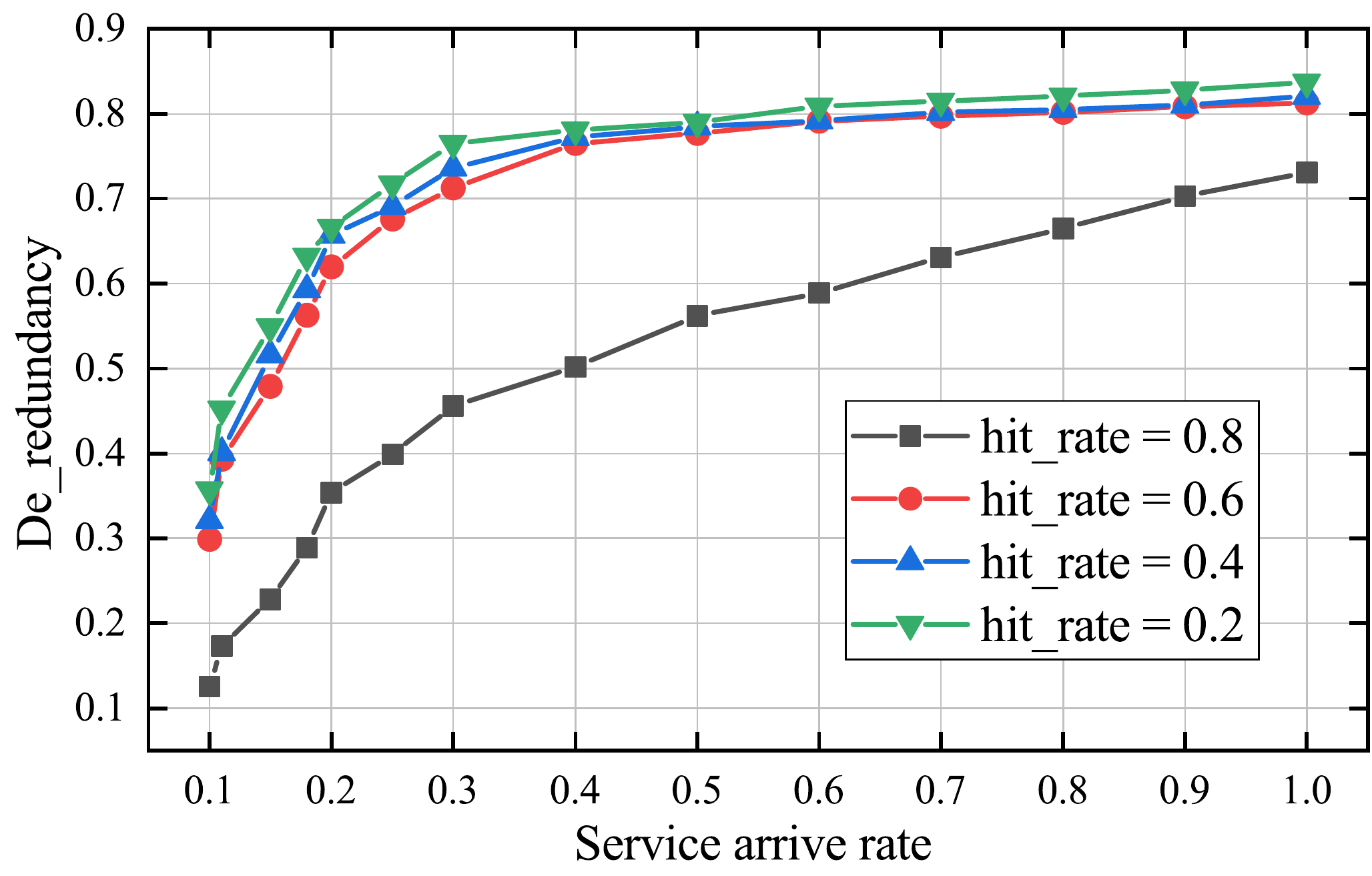}
  \caption{Deredundancy degree with the service arrival rate.}
  \label{de_redundancy arrive rate}
\end{figure}

Figure \ref{de_redundancy arrive rate} shows the relationship between the de-redundancy degree and service arrival rate under different requirements for the edge offloading rate. 
It can be seen from the figure that the higher the service arrival rate is, the more similar the services are in a short period of time, i.e., in time slot $\tau$, and the fewer services are retained and cached on the MEC servers after de-redundancy.
Under the requirement of an edge offloading rate of $0.2$, $0.4$, and $0.6$, the de-redundancy degree $\vartheta $ tends to be stable when the service arrival rate is greater than 0.6. 
This is because, to meet the requirements of edge computing in the network, the edge connectivity layer must cache a certain quantity of services. Therefore, with the increase in the service arrival rate, the de-redundancy degree $\vartheta $ in the service deployment stage will fluctuate upward only slightly, without a substantial increase.
In addition, the required edge offloading rate in the network will affect the de-redundancy degree of service deployment.
According to the longitudinal observations in Figure \ref{de_redundancy arrive rate}, the de-redundancy will decrease with the increase in the required edge offloading rate. In other words, the higher the required edge offloading rate is, the more services need to be cached on the MEC servers and the fewer services can be removed. This will lead to a lower degree of redundancy.

\begin{figure}[!t]
  \centering
  \includegraphics[width=2.5in]{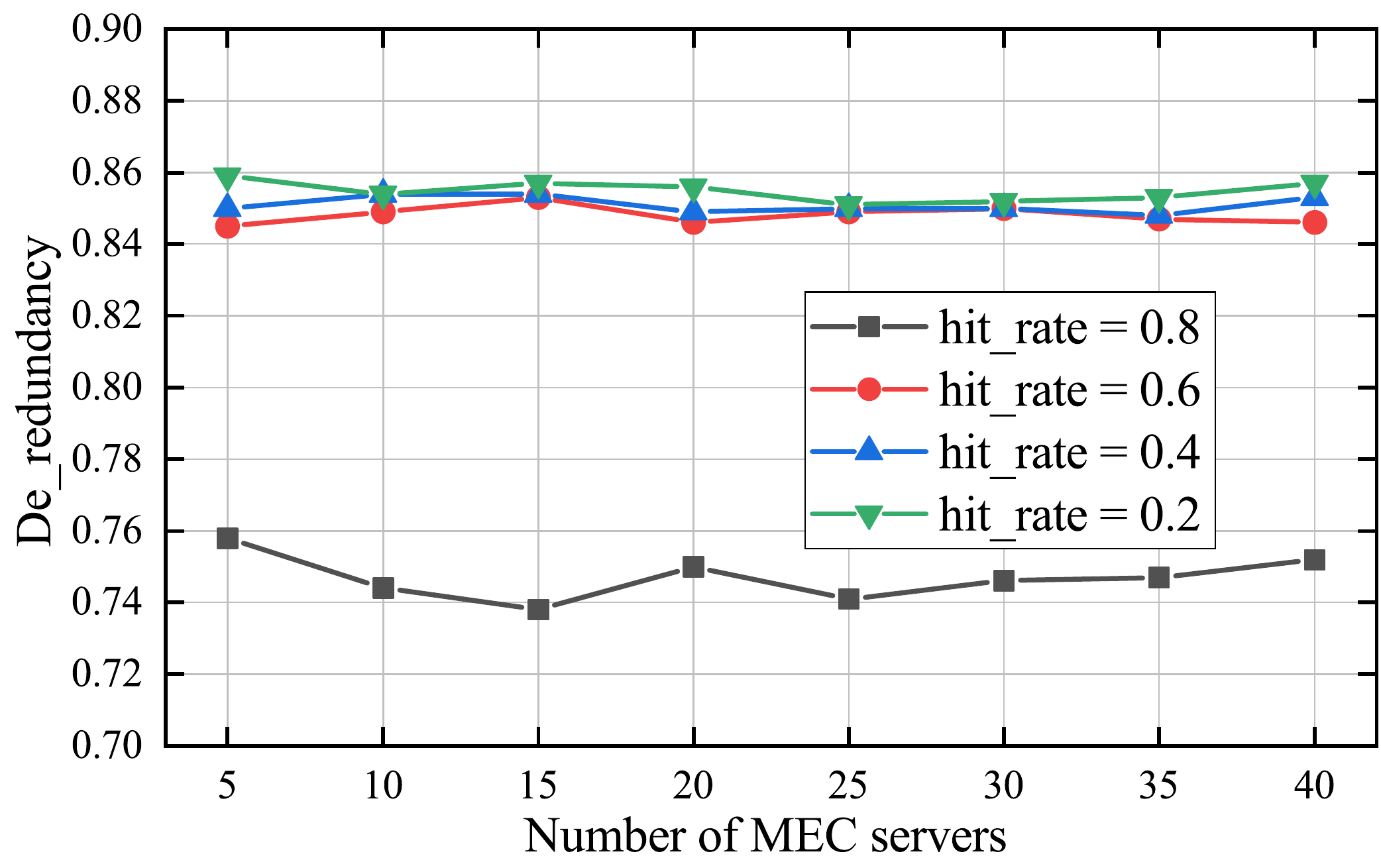}
  \caption{Deredundancy degree with the number of MEC servers.}
  \label{de_redundancy mec}
\end{figure}

As shown in Figure \ref{de_redundancy mec}, in the peak period (the service arrival rate $B_s$ is higher than 0.6), with the increase in the number of MEC servers, the de-redundancy degree remains basically stable.
Since the SD-AETO scheme models all MEC servers in the research scope as an AD-graph, the best service deployment scheme on the whole can be achieved regardless of the number of nodes.
When the requirement of the edge offloading rate is lower than $0.6$, the de-redundancy degree fluctuates up and down around $0.85$. 
When the edge offloading rate is higher than 0.6, similar to Figure \ref{de_redundancy arrive rate}, our scheme will sacrifice part of the service cache space to meet the demand of an excessive edge offloading rate. However, even so, this scheme can ensure a de-redundancy degree greater than $70\%$.
Combined with Figure \ref{de_redundancy arrive rate}, Figure \ref{de_redundancy mec} reflects the robustness and wide applicability of the SD-AETO scheme, and this scheme can meet the edge network requirements of peak or low peak periods at the same time.

\begin{figure}[!t]
  \centering
  \includegraphics[width=2.5in]{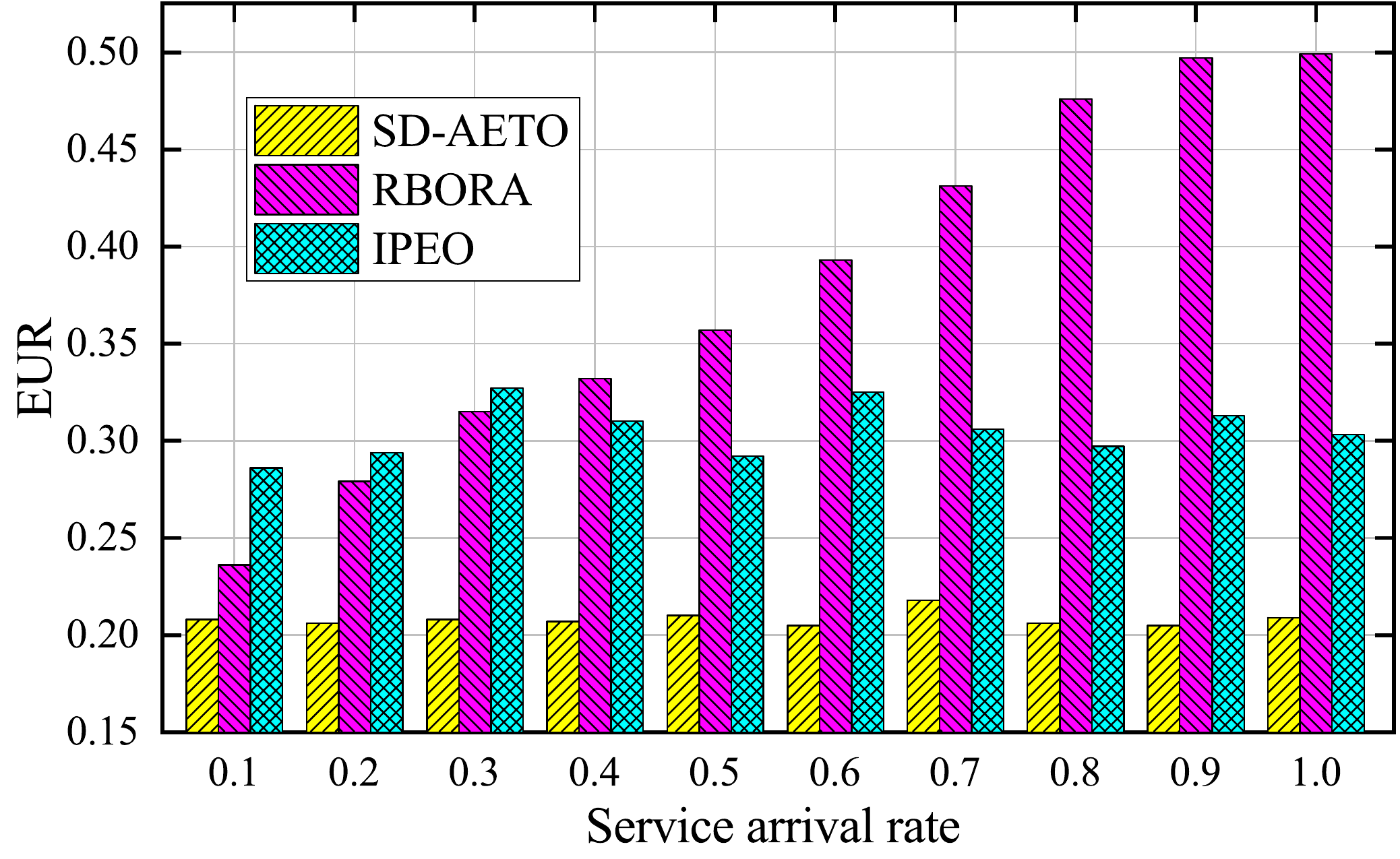}
  \caption{EUR with the service arrival rate.}
  \label{EURservice}
\end{figure}

Since the MEC server is not only used to provide task offloading services, we hope to minimize the utilization of resources on the MEC server in this paper so that the edge connectivity layer can provide more edge services for UEs.
Figure \ref{EURservice} shows the simulation results of the energy utilization ratio (EUR) of the SD-AETO scheme.
Compared with the other algorithms, our scheme has a low and stable EUR.
When the service peak period arrives, the edge resource occupation of RBORA \cite{RBORA} increases significantly, and the EUR of IPEO \cite{IPEO} hovers in the range $(0.27, 0.37)$, while our scheme only fluctuates slightly around $0.2$. This is because the proposed scheme can adapt to peak or low peak service periods simultaneously.

\begin{figure}[!t]
  \centering
  \includegraphics[width=2.5in]{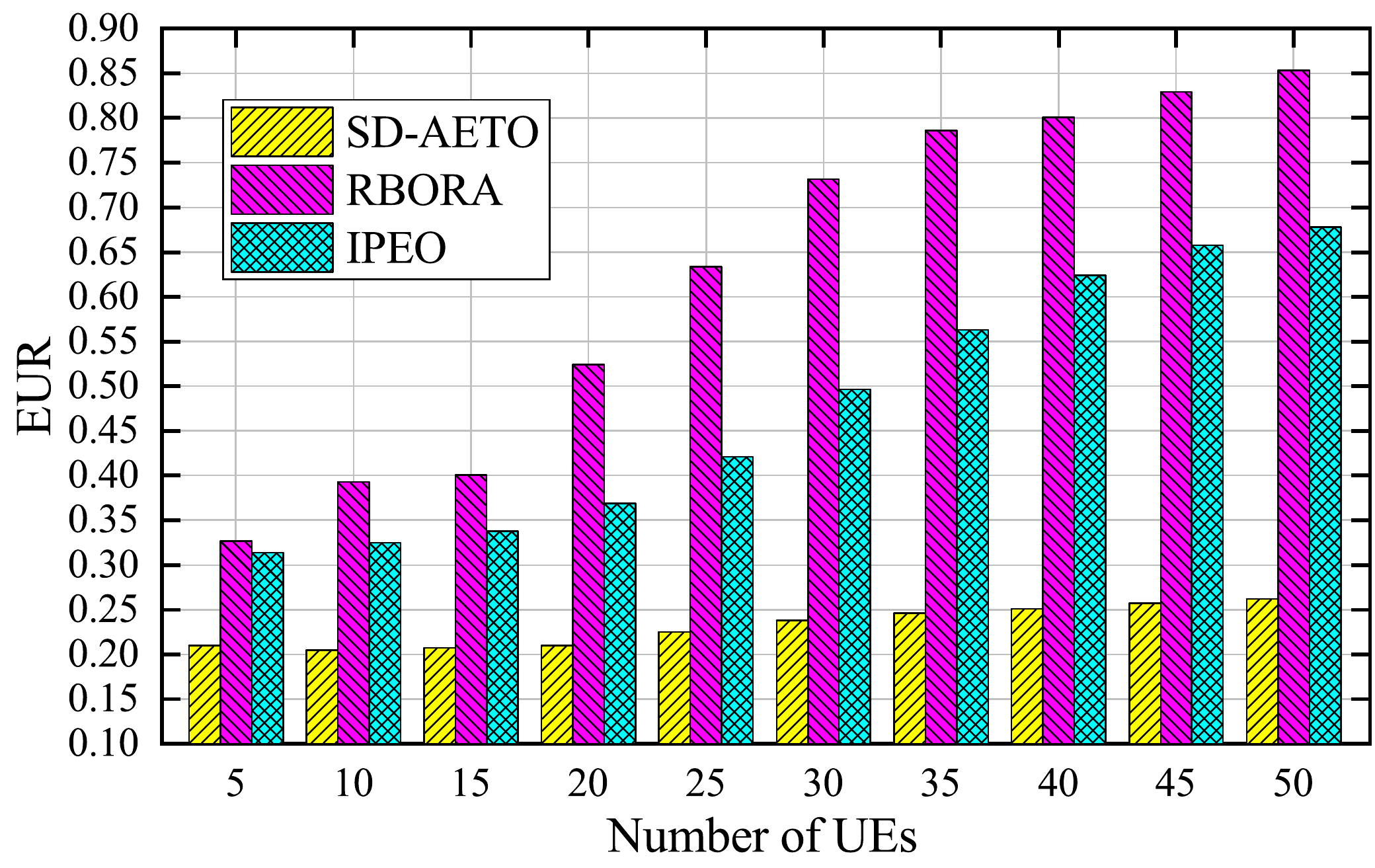}
  \caption{EUR with the number of UEs.}
  \label{EURUE}
\end{figure}

To further research the performance of the EUR of the SD-AETO scheme, Figure \ref{EURUE} gives the simulation curve with the number of UEs as the independent variable.
The EUR of our scheme remains at the lowest position. In addition, with the increase in UEs, the SD-AETO scheme has only a slight increase from $0.2$, while RBORA \cite{RBORA} and IPEO \cite{IPEO} have an obvious increase from $0.3$ so that most of the resources in the MEC servers are occupied in dealing with the offloading tasks in the end.
Therefore, we claim that compared with the other algorithms, the simulation results in Figure \ref{EURUE} intuitively reflect the stronger adaptability and superiority of the SD-AETO scheme in massive-task scenarios.

\begin{figure}[!t]
  \centering
  \subfloat[\small{Task arrival rate $B_u$.}]{\includegraphics[width=2.5in]{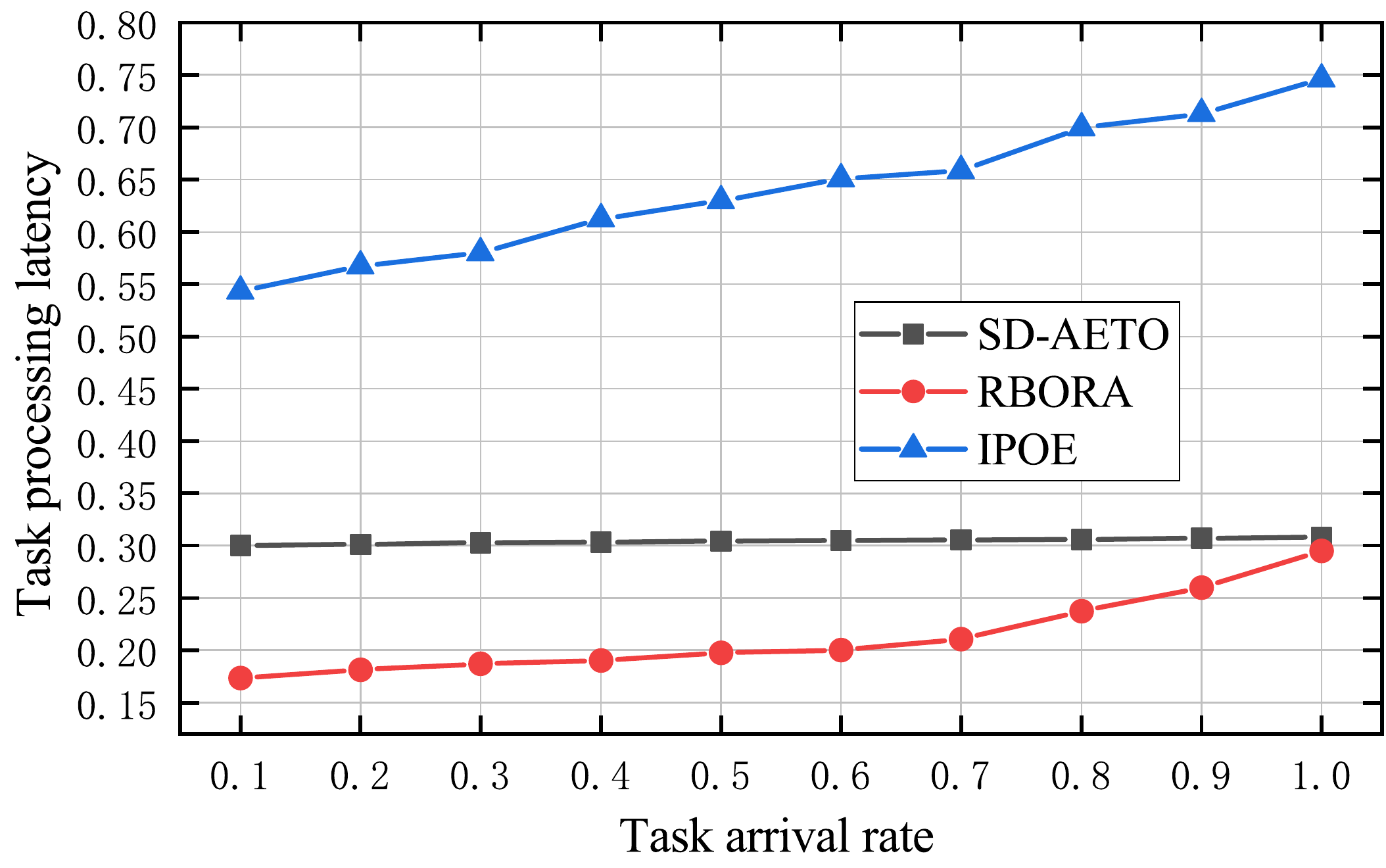}%
  \label{latency_task}}
  \\
  \subfloat[Number of UEs.]{\includegraphics[width=2.5in]{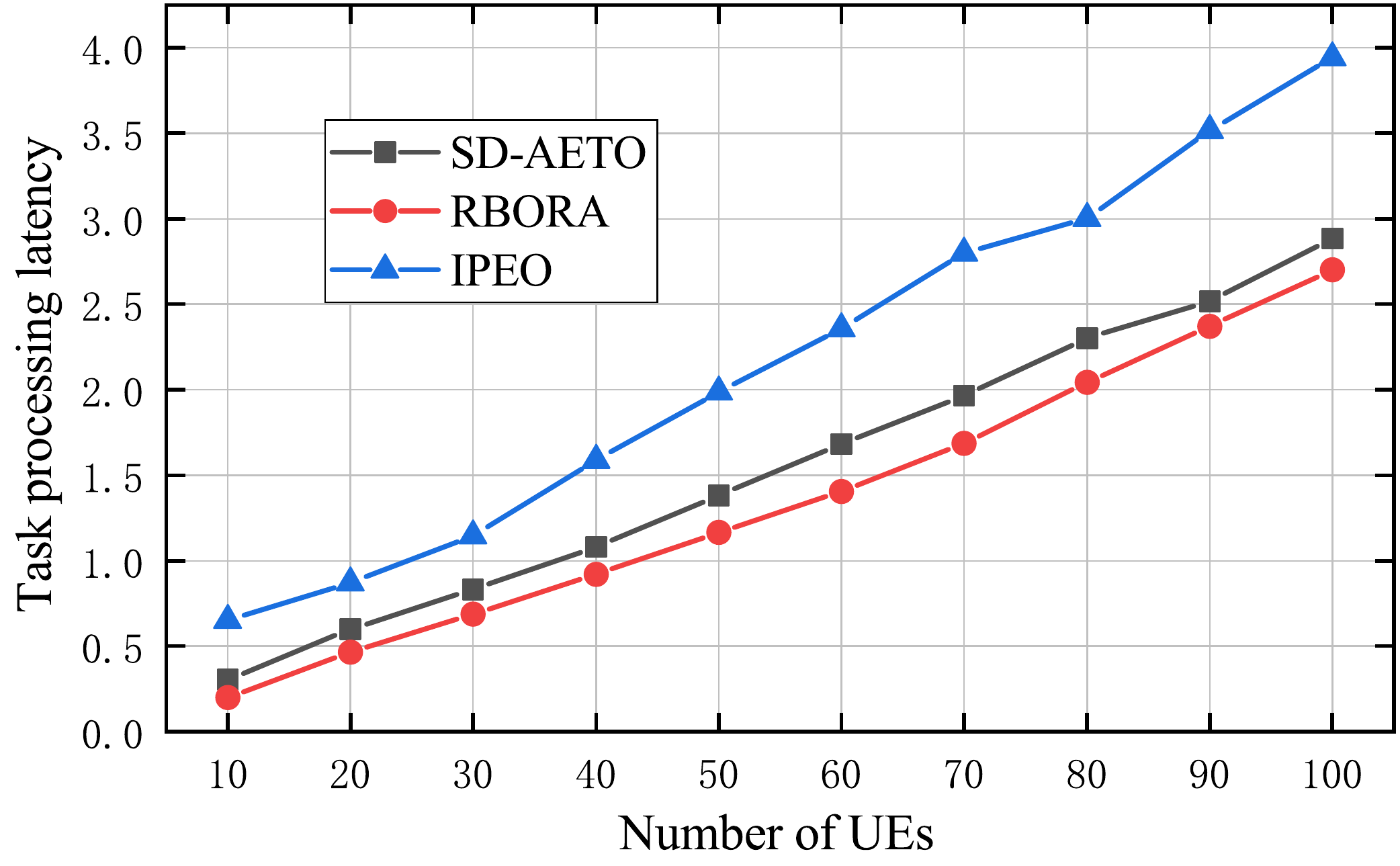}%
  \label{latency_UEs}}
  \caption{Task processing latency with task arrival rate and number of UEs.}
  \label{latency}
\end{figure}

To highlight the advantages of the proposed SD-AETO scheme, we also simulate the latency difference between the SD-AETO scheme and the comparison algorithms. It should be noted that the processing delay is determined according to the time at which the last task in the network is completely processed in time slot $\tau$. Figure \ref{latency} represents the task processing latency with the task arrival rate and number of UEs, where $latency(IPEO)>latency(SD-AETO)>latency(RBORA)$.
Comparing IPEO \cite{IPEO}, both our scheme and it involve the popularity of tasks or services and UEs, but the overall delay of our scheme is higher than that of RBORA \cite{RBORA}.
As shown in Figure \ref{latency_task}, with the increase in the task arrival rate in the network, the processing delay of the SD-AETO scheme is stable at approximately $0.3$ seconds, while IPEO \cite{IPEO} with the priority setting cannot maintain a low constant processing delay.
Comparing RBORA \cite{RBORA} without the priority setting, although the delay of our scheme is higher than that when the task arrival rate is low, the delay gap between them is less than $0.1$ in the peak period, and with the increase in the task arrival rate, the gap between them decreases until it can be ignored.
Figure \ref{latency_UEs} shows the trend of processing delay under the scenario of a changing number of UEs. The delay gap between our scheme and RBORA \cite{RBORA} is very small, which shows that although the SD-AETO scheme takes popularity into account, it does not significantly increase the processing delay of tasks. That is, from the perspective of UEs, they can obtain more optimized edge services while meeting their own offloading requirements.

\section{Conclusion}
To quantify the impact of popularities on services, we proposed a service popularity design method. Then, based on this popularity design method, service deployment was implemented in the edge connectivity layer. Furthermore, in the service deployment of the SD-AETO scheme proposed in this paper, we reconstructed the MEC servers and services as an AD-graph and solved the quota problem of the Steiner tree to find the optimal deployment strategy. Next, in the SD-AETO scheme implementation stage, we assigned priorities to each subtask based on the microservice popularity. Finally, we proposed an offloading queue design method to achieve the overall optimization goal of task offloading.
From the simulation results in section VI, it can be seen that the edge offloading rate of the proposed SD-AETO scheme is improved by $31\%$ compared with the other algorithms, and the EUR in the MEC servers is decreased by approximately $69\%$.


%





\ifCLASSOPTIONcaptionsoff
  \newpage
\fi



%



\bibliographystyle{IEEEtran}
\bibliography{_ref} 

%








\end{document}